\documentstyle[aps,rmp,psfig,multicol]{revtex}
\begin{document}
\title{Metallic behavior and related phenomena in two dimensions\vspace{5mm}}
\author{Elihu Abrahams}
\address{Center for Materials Theory, Physics Department, Rutgers University, Piscataway, New Jersey 08854}
\author{S.\ V.\ Kravchenko}
\address{Physics Department, Northeastern University, Boston,
Massachusetts 02115}
\author{M.\ P.\ Sarachik}
\address{Physics Department, City College of the City University of New
York, New York, New York 10031\vspace{2mm}}
\maketitle
\begin{abstract} For about twenty years, it has been the prevailing view that there can be no metallic state or metal-insulator transition in two dimensions in zero magnetic field.  In the last several years, however, unusual behavior suggestive of such a transition has been reported in a variety of dilute two-dimensional electron and hole systems.  The physics behind these observations is presently not understood.  We review and discuss the main experimental findings and suggested theoretical models.
\end{abstract}
\begin{multicols}{2}
\tableofcontents
\section{INTRODUCTION}
\label{sec:introduc} According to the scaling theory of localization (Abrahams, Anderson, Licciardello, and Ramakrishnan, 1979) there can be no metallic state in two dimensions in zero magnetic field ($B=0$).  Within this two decades-old theory, all carriers are localized in an infinitely large two-dimensional (2D) system at zero temperature.  With decreasing temperature the resistance is expected to grow logarithmically (``weak localization'') or exponentially (``strong localization''), becoming infinite as $T\rightarrow0$.  Although this prediction was made for 2D systems of non-interacting particles, subsequent theoretical work showed that weak interactions between the electrons increase the localization even further (Altshuler, Aronov, and Lee, 1980).  In the opposite limit of very strong interactions between particles, a 2D electron system is expected to become a Wigner crystal (see, {\it e.g.}, Tanatar and Ceperley, 1989); in the presence of even a small amount of disorder, such a crystal is expected to be pinned so that the system of crystallized electrons would not conduct at zero temperature.  Therefore, 2D systems were not expected to be conducting in either limit: weak (or absent), or very strong interactions between carriers.

Experiments performed in the early 1980's on different 2D systems confirmed these predictions.  Thin metallic films and silicon metal-oxide-semiconductor field-effect transistors (MOSFETs) displayed the expected logarithmic increase in resistivity (Dolan and Osheroff, 1979; Bishop~{\em et~al.}, 1980; Uren~{\em et~al.}, 1980).  At low electron densities, an exponential increase of the resistivity of silicon MOSFETs as a function of inverse temperature was reported (Uren~{\em et~al.}, 1980).  The agreement between theoretical expectations and experimental results was convincing, and for nearly two decades, the question of whether a conducting state is possible in 2D was considered resolved.

However, from time to time, indications appeared that the accepted view may not always be correct.  Finkelstein (1984) and Castellani~{\em et~al.} (1984) considered the interplay of disorder and interactions and showed that for weak disorder and sufficiently strong interactions, a 2D system scales toward a state with finite non-zero conductivity as temperature is lowered.  Unfortunately, the conclusion was not very definite since the theory's range of validity was exceeded as this ``metallic'' region was approached.  Therefore the possibility of a 2D metal was not seriously considered.  A number of experimental results also suggested that metallic behavior is possible in two dimensions.  From an analysis of experimental data obtained in GaAs/AlGaAs heterostructures, Gold (1991) concluded that a metal-insulator transition existed in clean samples. Pudalov~{\em et~al.} (1993b) and Shashkin~{\em et~al.} (1993,1994a,1994b) studied the phase diagram of the quantum Hall effect in low-disordered Si MOSFETs and GaAs/AlGaAs heterostructures and arrived at a similar conclusion.  In such 2D systems, in high magnetic field, each Landau level comprises a set of states with a spread in energy due to the presence of disorder.  The conventional view is that these states are all localized except for those at the center of the energy distribution, which are extended.  In the limit of zero magnetic field, these extended states are expected to float up indefinitely in energy (Khmelnitskii, 1984; Laughlin, 1984).  However, Pudalov~{\em et~al.} and Shashkin~{\em et~al.} found, contrary to expectations, that the extended states do not float up indefinitely in the limit of zero magnetic field, but coalesce instead at the Fermi level, thus allowing for a metallic state at $B=0$.  These observations were not well-recognized and the conventional wisdom persisted that there can be no metallic state in two dimensions in the absence of magnetic field.

In recent years, however, systematic studies of the temperature dependence of the resistance in zero magnetic field in a variety of dilute, low-disordered 2D systems have suggested that this point of view may be incorrect.  Metallic behavior (resistivity that decreases with decreasing temperature) has been observed down to the lowest accessible temperatures at electron ($n_s$) or hole ($p_s$) densities above some critical density $n_c$ (or $p_c$).  Below this critical density, the behavior of the resistance is insulating, thus suggesting that a metal-to-insulator transition in two dimensions occurs as the density is varied.  At the critical density, the resistivity is found to be nearly independent of temperature and of the order of the quantum unit of resistance, $h/e^2\approx25.6$~k$\Omega$.\footnote{Recall that in two dimensions, resistivity and resistance per square are the same quantity.}  Application of an external magnetic field of the order of a few Tesla, either parallel, tilted, or perpendicular to the 2D plane, suppresses the metallic behavior and gives rise to an enormous positive magnetoresistance on both sides of the transition.  Neither the metallic behavior nor its suppression by a magnetic field is currently understood.  The different models that have been suggested are briefly reviewed in Section~\ref{sec:explanation}.  In Section~\ref{sec:experiment}, we summarize the key experimental findings.  We concentrate mostly on dilute high-mobility Si MOSFETs, a system in which the unusual effects are particularly strong.  With few exceptions, we do not discuss effects at densities well above the critical density (Papadakis~{\em et al.}, 1999; Yaish~{\em et al.}, 2000; Dolgopolov and Gold, 2000), nor do we discuss experimental results on metal-insulator transitions in unorthodox systems like Ga[Al]As heterostructures with a layer of self-assembled quantum dots (Ribeiro {\it et al.}, 1999) in which the properties are quite different.

\section{EXPERIMENT}
\label{sec:experiment}
\subsection{Samples}

The first experiments that reported strong metallic temperature dependence of the resistivity and attributed the unusual behavior to the existence of a metallic state and a metal-insulator transition in 2D were performed on very low-disordered silicon MOSFETs (Kravchenko~{\em et al.}, 1994, 1995a).  Peak electron mobilities in these samples exceeded those in the samples used in previous studies by an order of magnitude, reaching more than $4\times10^4$~cm$^2$/Vs at $T=4.2$~K.  The very high quality of the samples allowed studies of the 2D system in a very dilute regime, i.e., at electron densities below $10^{11}$~cm$^{-2}$.  Instead of being small compared to the Fermi energy, the electron-electron interaction energy, $E_{\text{e-e}}$, is the dominant parameter at these low densities.  Estimates for Si MOSFETs at $n_s=10^{11}$~cm$^{-2}$ yield
\begin{equation}
E_{\text{e-e}}\sim\frac{e^2}{\epsilon}(\pi n_s)^{1/2}\approx10\text{ meV}
\end{equation}
while
\begin{equation}
E_F=\frac{\pi\hbar^2 n_s}{2m^*}\approx0.58\text{ meV}
\end{equation}
(where $e$ is the electron charge, $\epsilon$ is the dielectric constant, $E_F$ is the Fermi energy, and $m^*$ is the effective electron mass.  For a MOSFET in a (100) surface, a valley degeneracy of two is taken into account when calculating the Fermi energy.)  The dimensionless parameter $r_s\equiv E_{\text{e-e}}/E_F$ thus assumes values above 10 in these samples.  In the very dilute regime, 2D electrons are expected to form a Wigner crystal if the disorder is weak; a numerical simulation (Tanatar and Ceperley, 1989) predicted this should occur at $r_s\approx37\pm5$. In subsequent work, Chui and Tanatar (1995) showed that solidification should occur at even higher density when disorder is present. Therefore, it is reasonable to expect that the 2D system is a strongly correlated liquid at $r_s\sim10$.

Subsequent experiments in dilute silicon MOSFETs with different geometry and oxide thicknesses (Popovi\'{c}~{\em et~al.}, 1997) confirmed the earlier findings, and similar behavior was reported in a variety of other 2D systems, including p-SiGe heterostructures (Coleridge~{\em et~al.}, 1997; Lam~{\em et~al.}, 1997), p-GaAs/AlGaAs heterostructures (Hanein~{\em et~al.}, 1998a; Simmons~{\em et~al.}, 1998; Yoon~{\em et~al.}, 1999; Mills~{\em et~al.}, 1999), n-AlAs heterostructures (Papadakis and Shayegan, 1998),
\vbox{
\vspace{-1.5mm}
\hbox{
\psfig{file=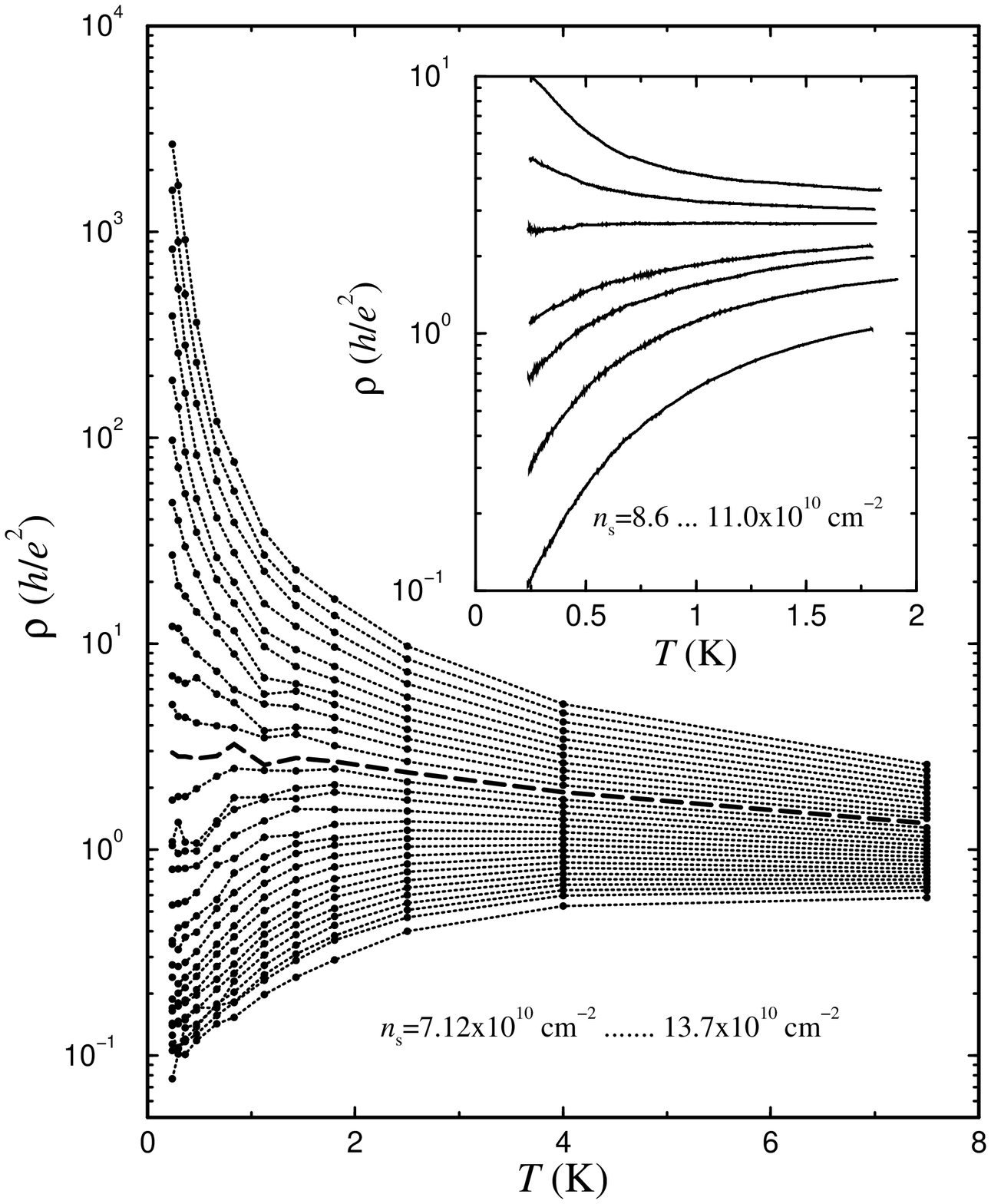,width=3.1in,bbllx=.5in,bblly=1.25in,bburx=7.25in,bbury=9.5in,angle=0}
}
\vspace{0.3in}
\hbox{
\hspace{-0.15in}
\refstepcounter{figure}
\parbox[b]{3.4in}{\baselineskip=12pt \egtrm FIG.~\thefigure.
Temperature dependence of the resistivity in a dilute low-disordered Si~MOSFET for 30 different electron densities (from Kravchenko~{\em et al.}, 1995a).  The inset shows accurate measurements of $\rho(T)$ close to the separatrix for another sample; the electron densities are 8.6, 8.8, 9.0, 9.3, 9.5, 9.9, and 11.0$\times$10$^{10}$~cm$^{-2}$ (from Sarachik and Kravchenko, 1999).\vspace{0.10in}
}
\label{fig:oklahoma_rt}
}
}
and n-GaAs/AlGaAs heterostructures (Hanein~{\em et~al.}, 1998c).  Values of $r_s$ in these studies varied from $\sim4$ (Hanein~{\em et~al.}, 1998c) to $>40$ (Yoon~{\em et~al.}, 1999; Mills~{\em et~al.}, 1999).

\subsection{Metallic and insulating behavior: Evidence for a metal-insulator transition}

The temperature dependence of the resistivity of a typical low-disordered Si MOSFET is plotted in Fig.~\ref{fig:oklahoma_rt}.  Data are shown for 30 different electron densities varying from $7.12\times10^{10}$ to $13.7\times10^{10}$~cm$^{-2}$; the corresponding values of $r_s$ vary between approximately 15 and 20.  At low electron densities (upper curves), the resistivity grows monotonically as the temperature decreases, behavior that is characteristic of an insulator.  For densities just below the thick long-dashed curve, henceforth referred to as the separatrix corresponding to the critical electron density $n_s=n_c$, the resistivity exhibits non-monotonic behavior: the resistance slowly increases with decreasing temperature for temperatures above a temperature $T^*\approx2$~K and decreases sharply at lower temperatures; thus the behavior is like that of an insulator for $T>T^*$ and like that of a metal for $T<T^*$.  At still higher $n_s$, the resistivity is almost constant at high temperatures and drops sharply at lower temperatures, displaying strong metallic dependence of temperature.  The separatrix between metallic and insulating behavior extrapolates to approximately $3h/e^2$ in the low-temperature limit for this MOSFET.  The detailed behavior of the resistivity in the immediate vicinity of the critical electron density is shown in the inset for another Si MOSFET.  The lowest curve shows a ten-fold drop in resistivity at $T<1.8$~K with no indication of any low-temperature saturation.  At a critical density $n_c=9.02\times10^{10}$~cm$^{-2}$, the resistivity is almost independent of temperature, and changes in density of only 3\% from $n_c$ cause strongly metallic or strongly insulating behavior.  The value of the resistivity at the separatrix is close to that in the main figure.

In Fig.~\ref{fig:hanein_rt}, the resistivity as a function of temperature is shown for a different dilute 2D system, holes in a GaAs/AlGaAs heterostructure.  The hole density varies between $8.9\times10^9$ and $6.4\times10^{10}$~cm$^{-2}$, corresponding to $r_s$ between approximately 9 and 24.  The main features of $\rho(T)$ are the same as in Si MOSFETs: the behavior of the resistivity is insulating at low hole densities, $p_s<p_c$ 
\vbox{
\vspace{6.0mm}
\hbox{
\psfig{file=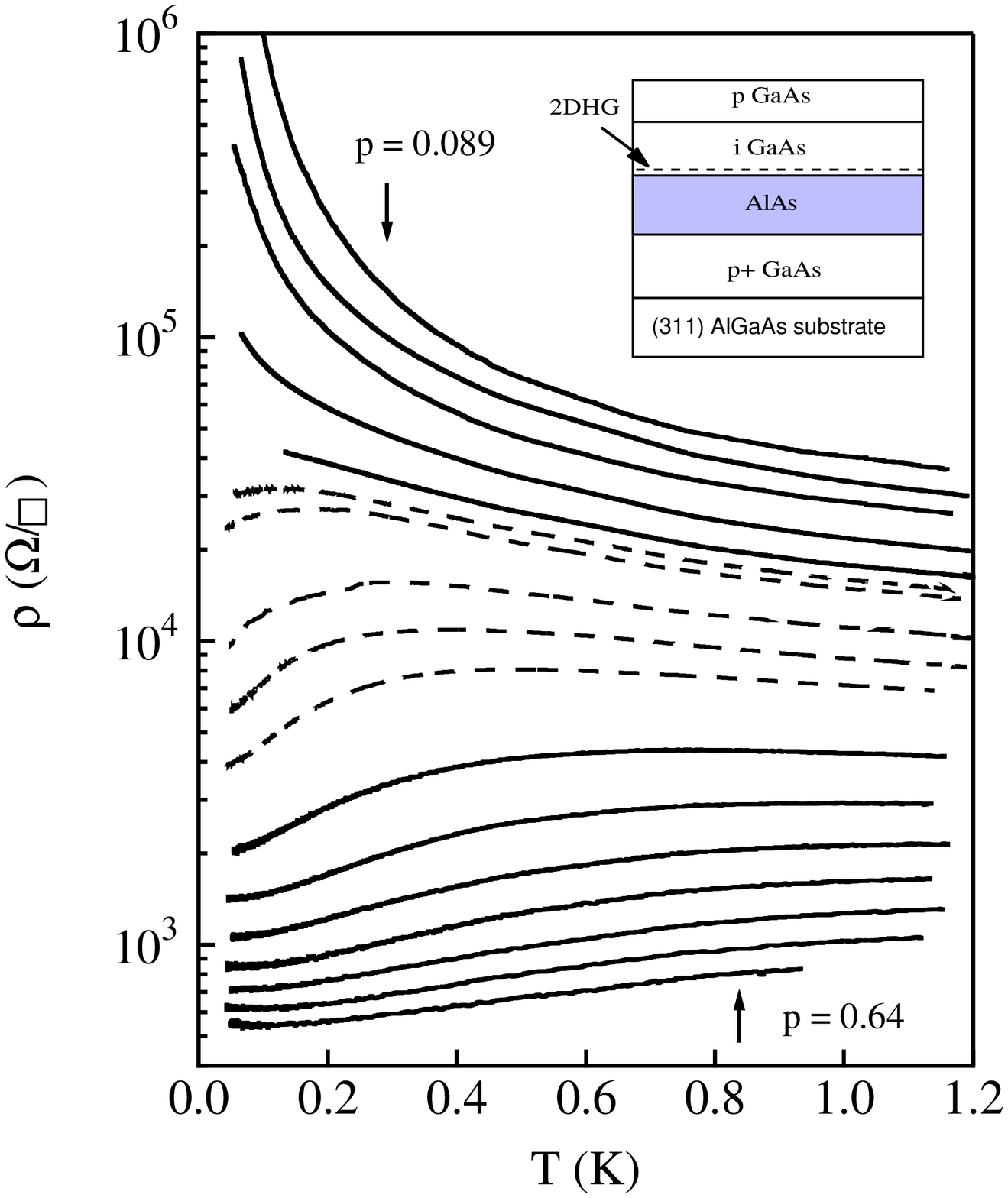,width=2.9in,bbllx=.5in,bblly=1.25in,bburx=7.25in,bbury=9.5in,angle=0}
}
\vspace{0.15in}
\hbox{
\hspace{-0.15in}
\refstepcounter{figure}
\parbox[b]{3.4in}{\baselineskip=12pt \egtrm FIG.~\thefigure.
For a 2D hole gas (2DHG) in p-GaAs/AlGaAs, resistivity per square as a function of temperature obtained at $B=0$ at various fixed hole densities, $p=$~0.089, 0.094, 0.099, 0.109, 0.119, 0.125, 0.130, 0.150, 0.170, 0.190, 0.250, 0.320, 0.380, 0.450, 0.510, 0.570, and 0.640$\times$10$^{11}$~cm$^{-2}$/Vs.  Note the three distinct regimes:  insulating regime at low densities, mixed regime at intermediate densities indicated by dashed lines, and a metallic-like regime at high densities.  Inset:  schematic presentation of a p-type ISIS (inverted semiconductor-insulator-semiconductor) structure used in the experiments.  From Hanein~{\em et al.} (1998a.)
}
\label{fig:hanein_rt}
}
}
(the upper solid curves); for $p_s\gtrsim p_c$, the resistivity shows insulating-like behavior at higher temperatures and drops by a factor of 2 to 3 at temperatures below a few hundred mK (the dashed curves); and at yet higher hole densities, the resistivity is metallic in the entire temperature range (the lower solid curves).  The curve which can approximately be identified as the separatrix between metallic and insulating behavior in the limit of low temperature extrapolates to approximately $1.5\,h/e^2$. We note that the range of carrier densities in Fig.~\ref{fig:hanein_rt} is much larger than that of Fig.~\ref{fig:oklahoma_rt}, and the apparent saturation of the resistivity at low temperatures, which is seen in Fig.~\ref{fig:hanein_rt} for the highest hole densities, is also observed in Si MOSFETs at electron densities higher than those shown in Fig.~\ref{fig:oklahoma_rt}.

Similar low-temperature drops in resistivity by a factor of 2 to 3 have been observed in several other dilute electron and hole systems: p-SiGe (Coleridge~{\em et~al.}, 1997), p-GaAs/AlGaAs (Yoon~{\em et~al.}, 1999; Mills~{\em et~al.}, 1999), and n-AlAs (Papadakis and Shayegan, 1998).  Qualitatively similar, but much weaker metallic temperature dependences of the resistivity were observed by Simmons~{\em et~al.} (1998) in p-GaAs/AlGaAs heterostructures and by Hanein~{\em et~al.} (1998c) in n-GaAs/AlGaAs.  In all these systems, the resistivity at the separatrix between metallic and insulating behavior, although not universal, is of the order of $h/e^2$.

\subsection{Experimental scaling}

It was found that the resistivity $\rho(T)$ of high-mobility silicon MOSFETs can be scaled with density and temperature over a range of temperature specified below.  As shown in Fig.~\ref{fig:scaling_t}, values of a scaling parameter $T_0(n_s)$ can be chosen (Kravchenko~{\em et al.}, 1995a), one for each density $n_s$, that yield a collapse of the data onto two curves: an insulating branch for densities $n_s<n_c$ and a metallic branch for $n_s>n_c$.  The resistivity is given by:
\begin{equation}
\rho(T,n_s)= \rho_c f_1[T/T_0(n_s)],
\end{equation}
where $\rho_c$ is the value of the resistivity at the critical density.  The scaling breaks down above $T^*$, and at very low temperatures, where the resistance has a much weaker $T$ dependence.  Nonetheless, it is noteworthy that a satisfactory overlap is obtained for more than ten curves on each side of the transition over a temperature range from 
\vbox{
\vspace{-1mm}
\hbox{
\hspace{3mm}
\psfig{file=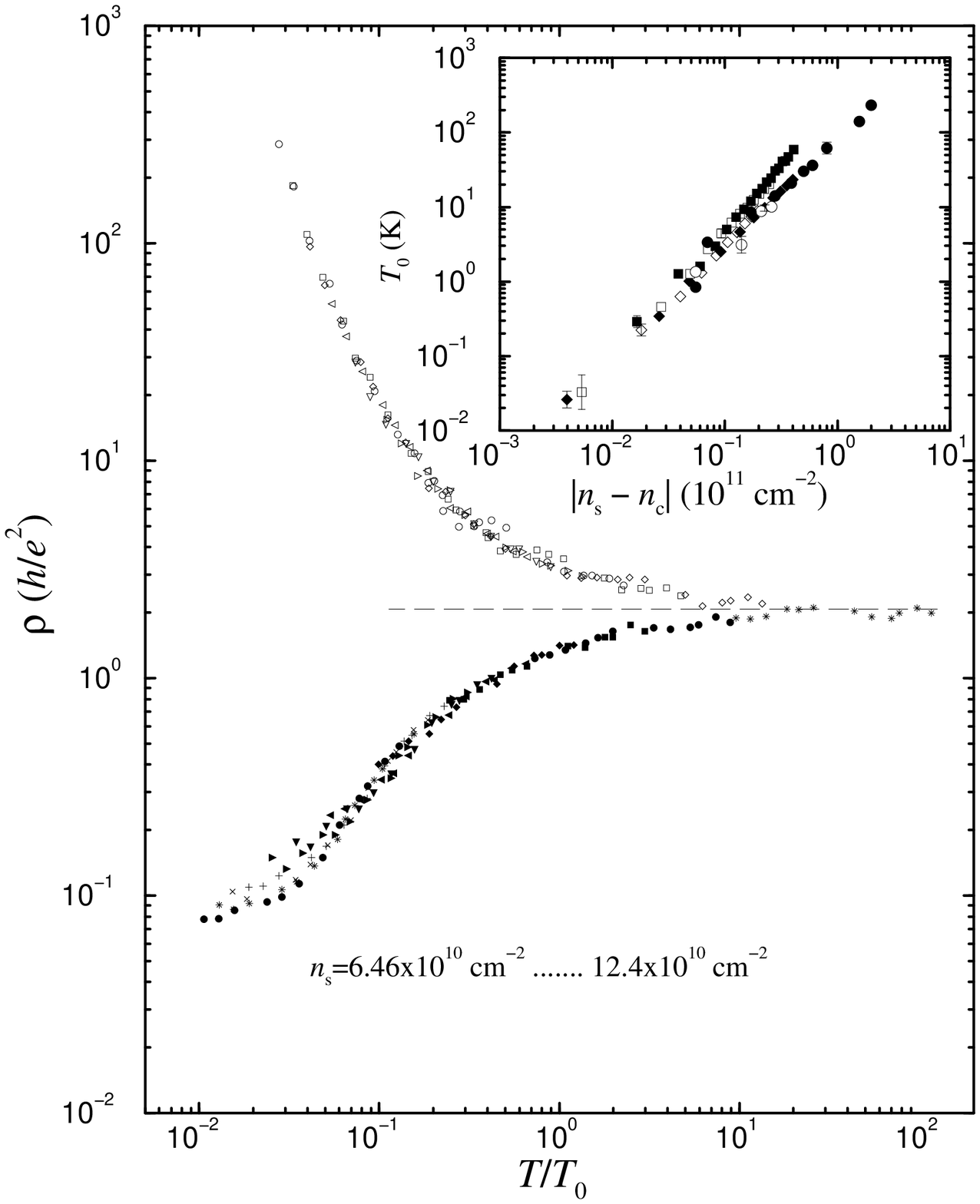,width=2.9in,bbllx=.5in,bblly=1.25in,bburx=7.25in,bbury=9.5in,angle=0}
}
\vspace{0.15in}
\hbox{
\hspace{-0.15in}
\refstepcounter{figure}
\parbox[b]{3.4in}{\baselineskip=12pt \egtrm FIG.~\thefigure.
For a silicon MOSFET, resistivity versus $T/T_0$, with $T_0$'s chosen to yield scaling with temperature.  The inset shows the scaling parameter, $T_0$, versus deviation from the critical point, $|n_s-n_c|$; data are shown for silicon MOSFETs obtained from three different wafers.  Open symbols correspond to the insulating side and closed symbols to the metallic side of the transition.  From Kravchenko~{\em et al.} (1995a).\vspace{0.10in}
}
\label{fig:scaling_t}
}
}
approximately 0.25~K to 2~K where the resistivity of each curve changes by an order of magnitude.

It is remarkable that $T_0(n_s)$ is independent of the sign of $\delta_n\equiv(n_s-n_c)/n_s$.  The inset in Figure~\ref{fig:scaling_t} shows $T_0$ as a function of the absolute deviation from the critical point, $|n_s-n_c|$, on a log-log scale for both metallic and insulating curves and for three different samples.  It is important to note that the dependence is a power law, $T_0\propto|\delta_{n}|^b$, with approximately the same power $b=1.60\pm0.1$ for all three samples and for both metallic and insulating curves.  In addition, it was observed that the metallic and insulating curves are reflection symmetric in the temperature range above 300~mK and below $T^*$ (Simonian~{\em et al.}, 1997a) as can be seen in Fig.~\ref{fig:scaling_t}.\footnote{Similar symmetry had been previously reported by Shahar~{\em et al.} (1996, 1997) near the quantum Hall effect to insulator transition, where it was attributed to charge-flux duality.}  Thus, the normalized scaled resistivity on either side of the transition is symmetric with its inverse on the other side: $\rho^*(n_s-n_c,T)=1/\rho^*(n_c-n_s,T)$ (here $\rho^*\equiv\rho/\rho_c$).  Similar symmetry was also reported by Popovi\'{c}~{\em et al.} (1997) and Simmons~{\em et al.} (1998).  Dobrosavljevi{\'c}~{\em et~al.} (1997) showed that the observed scaling and reflection symmetry are consequences of a simple analysis assuming that a $T=0$ quantum critical point describes the metal-insulator transition.  Within quantum critical scaling the power law exponent $b$ in $T_0\propto|\delta_{n}|^b$ is given by $b=z\nu_1$, where $z$ is the dynamical exponent and $\nu_1$ is the correlation length exponent (Sondhi~{\em et~al.}, 1997).\footnote{We use the symbol $\nu_1$ for the correlation length exponent to distinguish it from $\nu$, the Landau level filling factor.}  A discussion of the metal-insulator transition as an example of a quantum critical point has been given by Abrahams and Kotliar (1996).

Scaling analyses have been applied to other 2D systems with varying degrees of success.  While a number of experiments have yielded scaling exponents $b=z\nu_1$ between 1.25 and 1.6 for Si MOSFETs (Kravchenko~{\em et~al.}, 1994, 1995a, 1996; Popovi\'{c}~{\em et~al.}, 1997) and p-SiGe heterostructures (Coleridge~{\em et~al.}, 1997), attempts to scale the resistivity of p-GaAs/AlGaAs heterostructures have yielded much larger exponents (Simmons~{\em et~al.}, 1998), or have failed entirely (Hanein~{\em et~al.}, 1998a; Yoon~{\em et~al.}, 1999).  In general, it appears that scaling yields large exponents or breaks down for systems which exhibit relatively weak metallic behavior, as in some p-GaAs/AlGaAs heterostructures and low-mobility MOSFETs.  On the other hand, the more dramatic the metallic behavior the better the scaling fits, as in high-quality Si MOSFETs.

\subsection{Temperature and density dependence of resistivity}

The temperature dependence of the resistivity on opposite sides of $n_c$ ceases to be symmetric as one moves away from the transition. This is not inconsistent with scaling (Dobrosavljevi{\'c}~{\em et~al.}, 1997).  On the insulating side, the resistance was found (Mason~{\em et~al.}, 1995) to obey
\begin{equation}
\rho(T)=\rho_0\text{ exp}(T_1/T)^{1/2}.
\end{equation}
This is the form associated with variable-range hopping between localized states under the influence of the Coulomb interaction, commonly referred to as Efros-Shklovskii hopping (Efros and Shklovskii, 1975).  According to the theory, the prefactor $\rho_0$ is expected to be a weak function of temperature.  In contrast, Mason~{\em et~al.} (1995) found $\rho_0$ to be temperature-independent and close to the quantum unit of resistance, $h/e^2$.  This unexpected behavior --- Efros-Shklovskii hopping with a constant prefactor $\rho_0\approx h/e^2$ --- was also found recently by Khondaker~{\em et~al.} (1999) in insulating $\delta$-doped GaAs/AlGaAs heterostructures.

The temperature dependence of the resistivity has not been definitively established on the metallic side of the 
\vbox{
\vspace{-12mm}
\hbox{
\hspace{-0.20in}
\psfig{file=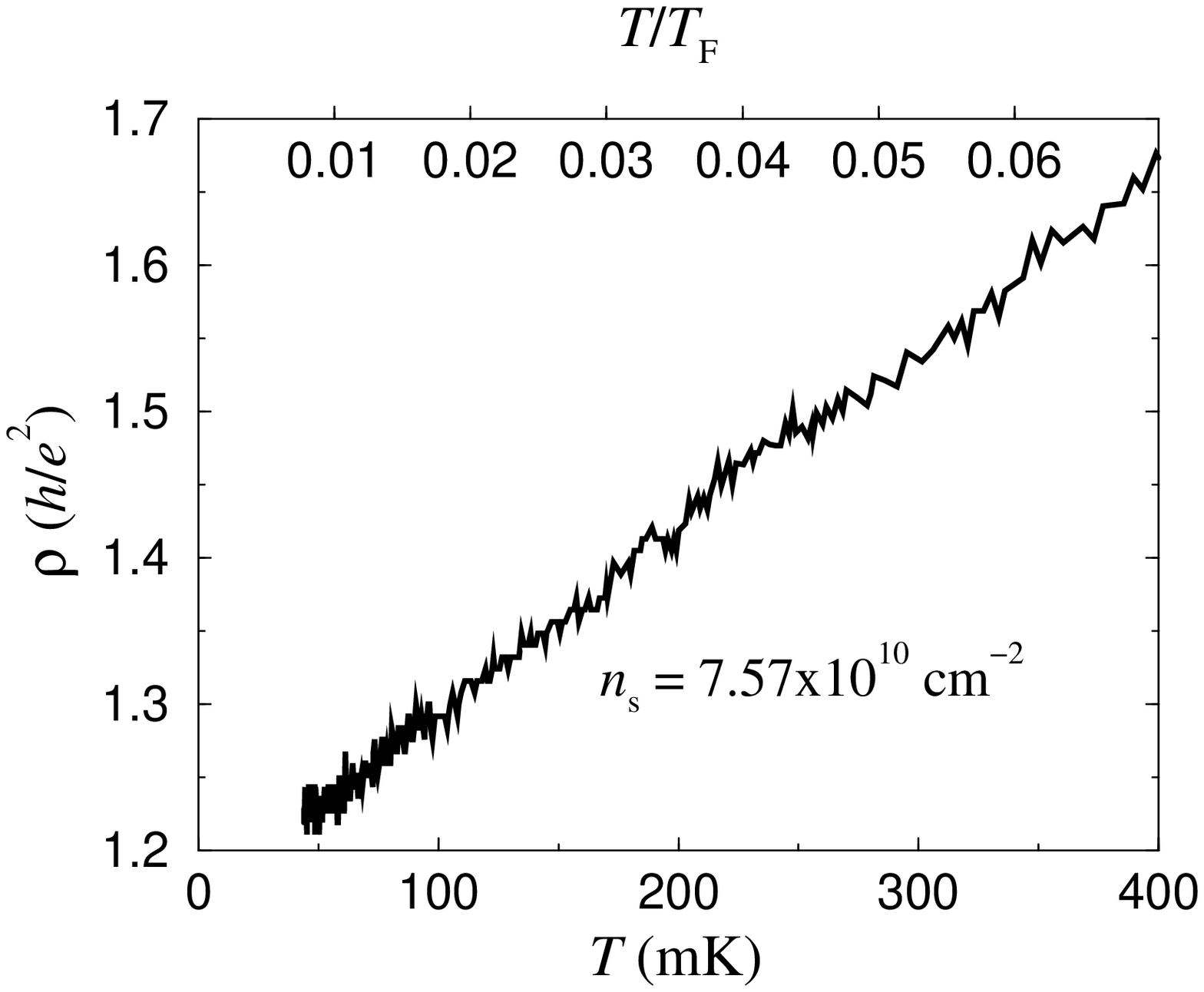,width=3.7in,bbllx=.5in,bblly=1.25in,bburx=7.25in,bbury=9.5in,angle=-90}
}
\vspace{0.1in}
\hbox{
\hspace{-0.15in}
\refstepcounter{figure}
\parbox[b]{3.4in}{\baselineskip=12pt \egtrm FIG.~\thefigure.
Resistivity of a ``just-metallic'' Si MOSFET as a function of temperature (lower x-axis) and as a function of the ratio $T/T_F$ (upper x-axis).  From Kravchenko and Klapwijk (2000).
\vspace{0.10in}
}
\label{fig:neu_rt}
}
}
critical carrier density.  It was first suggested by Pudalov (1997b) that for electron densities where the resistivity is a monotonically increasing function of the temperature, it can be approximated by the expression:
\begin{equation}
\rho(T)=\rho_1+\rho_2\, \text{exp}\,
\left[-(T_2/T)^{\gamma}\right],\label{Eq:exp}
\end{equation}
where $\rho_1$, $\rho_2$, and $T_2$ are temperature-independent (but $n_s$-dependent) parameters, and $\gamma$ is a constant of the order of 1.  This form is inconsistent with the resistivity-conductivity symmetry described above.  However, it should be noted that the symmetry is expected to hold only within the ``quantum critical region" close to the critical density, and is not expected to be valid below some temperature.

The activated scattering form, Eq.\ref{Eq:exp}, provides an adequate fit with $\gamma=1$ to data for some systems (Hanein~{\em et~al.}, 1998a; Papadakis and Shayegan, 1998; Mills~{\em et~al.}, 1999). In p-SiGe heterostructures (Coleridge~{\em et~al.}, 1997), where the resistance changes more gradually with temperature, $\gamma\approx0.5$ gives much better results.  The ratio $\rho_2/\rho_1$ decreases with electron density (Papadakis and Shayegan, 1998; Mills~{\em et~al.}, 1999), reflecting the fact that the metallic drop is relatively stronger at low carrier densities and weakens as the density increases.  At higher densities only the temperature dependence due to Bloch-Gruneisen phonon scattering is seen (Mills~{\em et~al.}, 1999).

The ultimate, zero-temperature fate of the resistivity for densities $n_s>n_c$ is an issue of central importance which has yet to be resolved.  We first consider high-mobility Si MOSFETs, a system where the anomalous effects are particularly strong.  At the critical electron density ({\em i.e.}, at the separatrix) in a high-quality sample, the resistivity is essentially independent of temperature from approximately 1~K down to about 35~mK, the lowest temperature reached in that experiment (Kravchenko and Klapwijk, 2000).  Just below the separatrix (for ``just-metallic'' electron densities, $n_s\gtrsim n_c$), the resistivity in MOSFETs decreases with decreasing temperature below $T^*$ down to the lowest accessed temperatures with no indication of an increase in resistivity at very low temperatures.  Note that the temperature reached in the experiment shown in Fig.~\ref{fig:neu_rt} is less than 1\% of the Fermi temperature, $T_F$ (see the upper axis on the graph) which is about 5~K in this experiment.  Thus, the electron system is definitely in the degenerate regime.  We emphasize that the metallic temperature dependence clearly overpowers quantum localization, which is expected to be very strong at resistivities of the order of $h/e^2$ and higher, where $k_F\ell\sim1$, and at temperatures $T\ll T_F$ (here $k_F$ is the Fermi wave number and $\ell$ is the mean free path).

At higher carrier density, the metallic resistivity varies much more slowly with temperature and it should not be examined on a log$\,\rho$ vs $T$ plot.  In this regime, it was found (Pudalov~{\em et~al.}, 1998b, 1999a) to be metallic-like up to $n_s\approx40\times n_c$ and reminiscent of weak localization (insulating-like) at still larger $n_s$.  This implies that $\rho_1$ in Eq.\ref{Eq:exp} may be weakly temperature-dependent rather than constant.  Of course, at such high densities the electron-electron interactions are no longer dominant and one expects to reenter the regime of weak localization as seen, for example, by Bishop~{\em et al.} (1982) in MOSFETs with very high electron densities around $10^{12}$ cm$^{-2}$.  This has been confirmed by recent measurements in both p-GaAs/AlGaAs (Hamilton~{\em et al.}, 1999) and Si MOSFETs (Pudalov~{\em et~al.}, 1999a).

To close this section, we consider the behavior in p-GaAs/AlGaAs and p-SiGe heterostructures.  In those heterostructures where the metallic effects are most prominent (see, {\it e.g.}, Coleridge~{\em et al.}, 1997; Hanein~{\em et~al.}, 1998a; Yoon~{\em et~al.}, 1999; Mills~{\em et~al.}, 1999), the behavior of the resistivity, while not as dramatic on the metallic side, is similar to that in Si MOSFETs: it is insulating-like at $n_s<n_c$, temperature-independent at the separatrix (Hanein~{\em et~al.}, 1998b), and metallic down to the lowest accessed temperatures at $n_s>n_c$.  However, in 2D systems where the low-temperature drop of the resistivity is weak, typically a few percent (Hamilton~{\em et~al.}, 1999; Simmons~{\em et~al.}, 2000) rather than factors of two or more, the behavior near the critical carrier density is qualitatively different.  The curves that look metallic at higher temperatures show a low-temperature upturn of the resistivity signaling a reentry into the insulating phase at sufficiently low temperatures (Simmons~{\em et~al.}, 2000; Senz~{\em et~al.}, 2000); there is no temperature-independent separatrix.  The behavior is similar to that seen in low-mobility Si MOSFETs or in MOSFETs with local magnetic moments (Feng {\it et al.}, 1999).  In these samples there is no sign of a metal-insulator transition.

\subsection{Nonlinear effects}

All the resistivity data discussed so far were obtained in the linear regime, {\em i.e.}, in the limit of zero electric field, $E\rightarrow0$.  The $I-V$ curves become non-linear when the electric field energy exceeds the thermal energy, $k_BT$, and the resistivity is then a function of electric field.  General arguments (Sondhi~{\em et~al.}, 1997) show that in the quantum critical region of a zero-temperature critical point (metal-insulator transition) if the resistivity scales with temperature it should also scale with electric field as:
\begin{equation}
\rho\,(E,n_s)=\rho_cf_2\,(\delta_n/E^{1/a}).
\end{equation}
Recall the exponent $b=z\nu_1$ for scaling with temperature; for electric field scaling, the exponent is $a=(z+1)\nu_1$.  Combining the electric field data with the temperature data allows separate determinations of the exponents $z$ and $\nu_1$, as was done for the superconducting transition in thin disordered films (Yazdani and Kapitulnik, 1995).

The resistivity of Si MOSFETs was determined as a function of electric field $E$ for different electron densities using $\rho=(V/I)\cdot(W/L)$ ($W$ is the sample width and $L$ is the distance between potential contacts) and the electric field $E=V/L$.  The curves shown in Fig.~\ref{fig:re} as a function of electric field are quite 
\vbox{
\vspace{4mm}
\hbox{
\hspace{3mm}
\psfig{file=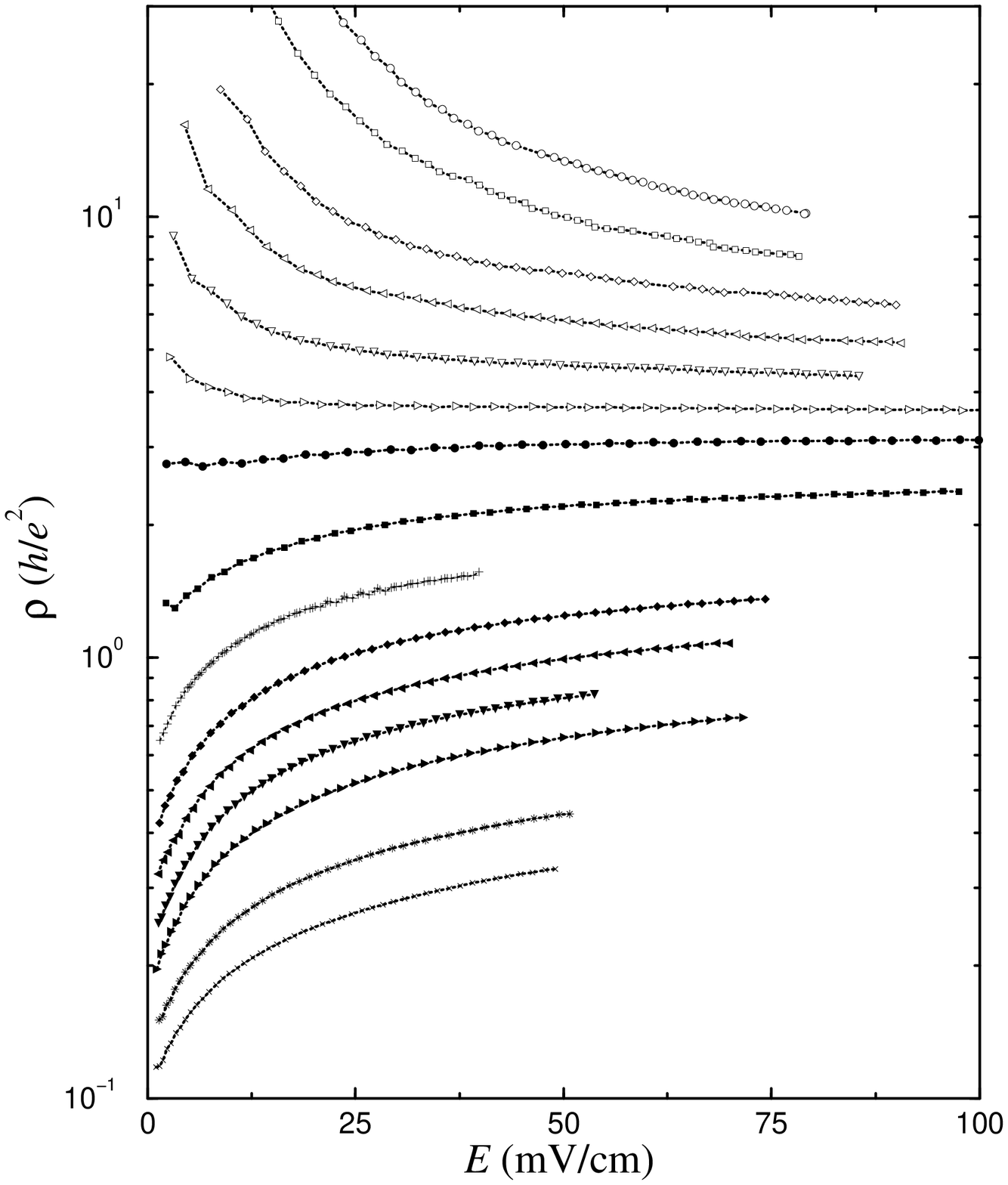,width=2.8in,bbllx=.5in,bblly=1.25in,bburx=7.25in,bbury=9.5in,angle=0}
}
\vspace{0.15in}
\hbox{
\hspace{-0.15in}
\refstepcounter{figure}
\parbox[b]{3.4in}{\baselineskip=12pt \egtrm FIG.~\thefigure.
Resistivity of a silicon MOSFET as a function of electric field for electron densities $n_s=$~7.81, 7.92, 8.03, 8.14, 8.25, 8.36, 8.47, 8.70, 8.91, 9.13, 9.35, 9.57, 9.79, 10.34, and 10.78$\times$10$^{10}$~cm$^{-2}$ at $T=0.22$~K.  From Kravchenko~{\em et~al.} (1996).
}
\label{fig:re}
}
}
similar to the curves of resistivity, $\rho(T)$, as a function of temperature shown earlier: the system is insulating ($d\rho/dE<0$) below some electron density $n_s<n_c$, metallic ($d\rho/dE>0$) above this density, and there is a well-defined flat separatrix between metallic and insulating behavior.  However, there is an essential difference between $\rho(E)$ and $\rho(T)$ for the ``just-metallic'' curves ($n_s\gtrsim n_c$): the $\rho(E)$ dependence is always monotonic while $\rho(T)$ displays a maximum at $T=T^*$ as discussed earlier.  A similar difference between $\rho(E)$ and $\rho(T)$ was observed in p-GaAs/AlGaAs heterostructures by Yoon~{\em et al.} (1999).
Plotting the resistivity against the scaling variable, $|\delta_n|/E^{1/a}$, yields a collapse of all the data onto two distinct branches, as shown in Fig.~\ref{fig:scaling_e}.  (Data for $E\rightarrow0$, where thermal rather than electric field energies dominate, are not included.)  The quality of the scaling with electric field is considerably higher than that for temperature scaling.  Combining the two yields a correlation length exponent $\nu_1$ between 1.5 and 1.9, and dynamical exponents $z$ between 0.8 and 1.2 in Si MOSFETs (Kravchenko~{\em et al.}, 1996; Heemskerk and Klapwijk, 1998) and $z=1$ in p-SiGe heterostructures (Senz~{\em et~al.}, 1999).  A microscopic basis for electric field scaling has been proposed by Leadbeater~{\em et~al.} (1999).  We note that the exponent $z$ has also been found to be close to 1 in other interacting 2D systems; see, {\it e.g.}, Yazdani and Kapitulnik (1995) for the case of the superconducting transition in thin films and Wei~{\em et~al.} (1994) for the transition between two neighboring QHE plateaus.

\vbox{
\vspace{-9.7mm}
\hbox{
\hspace{3mm}
\psfig{file=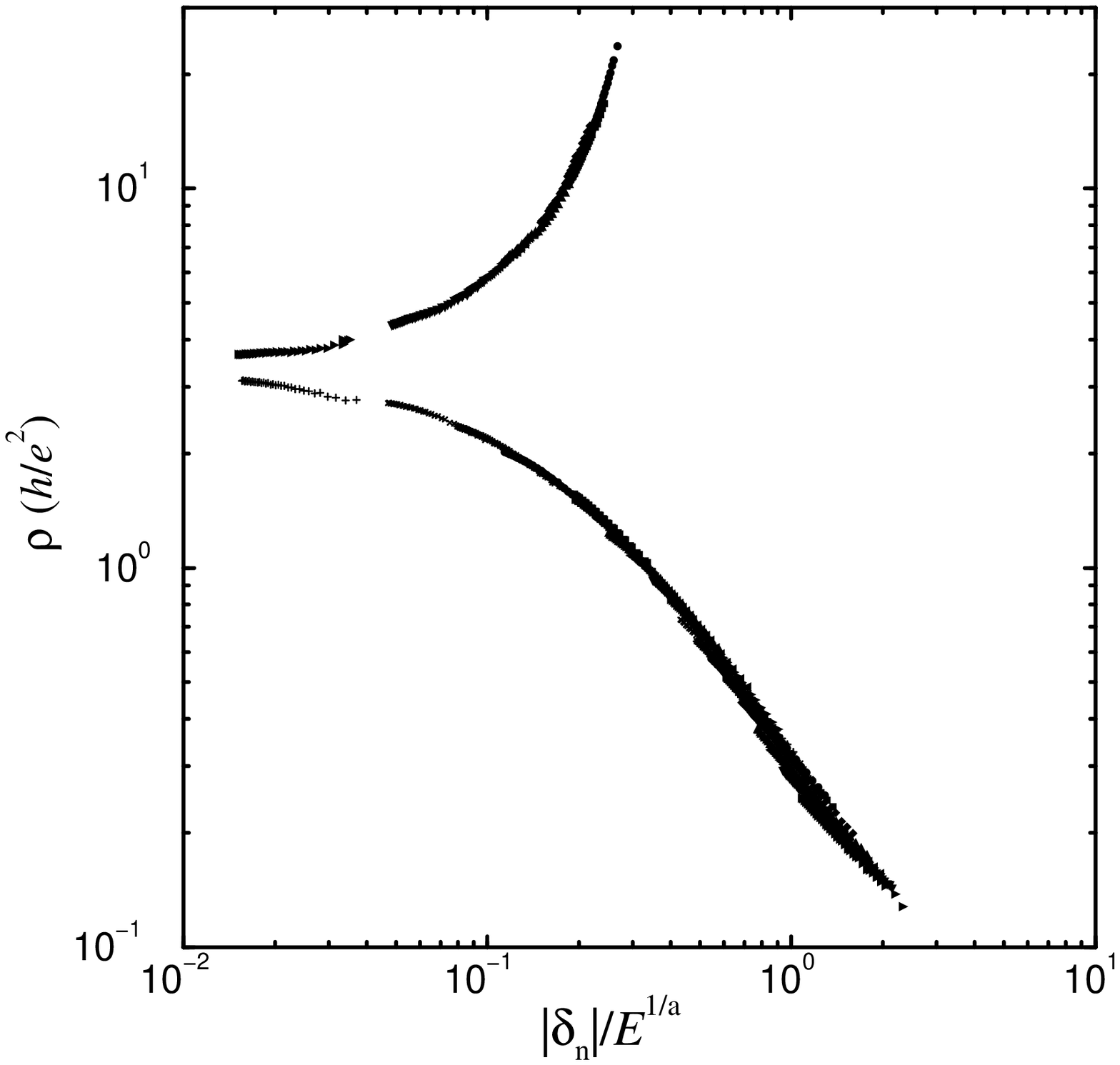,width=2.8in,bbllx=.5in,bblly=1.25in,bburx=7.25in,bbury=9.5in,angle=0}
}
\vspace{0.5in}
\hbox{
\hspace{-0.15in}
\refstepcounter{figure}
\parbox[b]{3.4in}{\baselineskip=12pt \egtrm FIG.~\thefigure.
Demonstrating scaling with electric field, the resistivity of a silicon MOSFET at 0.22~K is plotted as a function of $|\delta_n|/E^{1/a}$ for $a=2.7$.  From Kravchenko~{\em et al.} (1996).\vspace{0.10in}
}
\label{fig:scaling_e}
}
}

\vbox{
\vspace{6.5mm}
\hbox{
\hspace{7mm}
\psfig{file=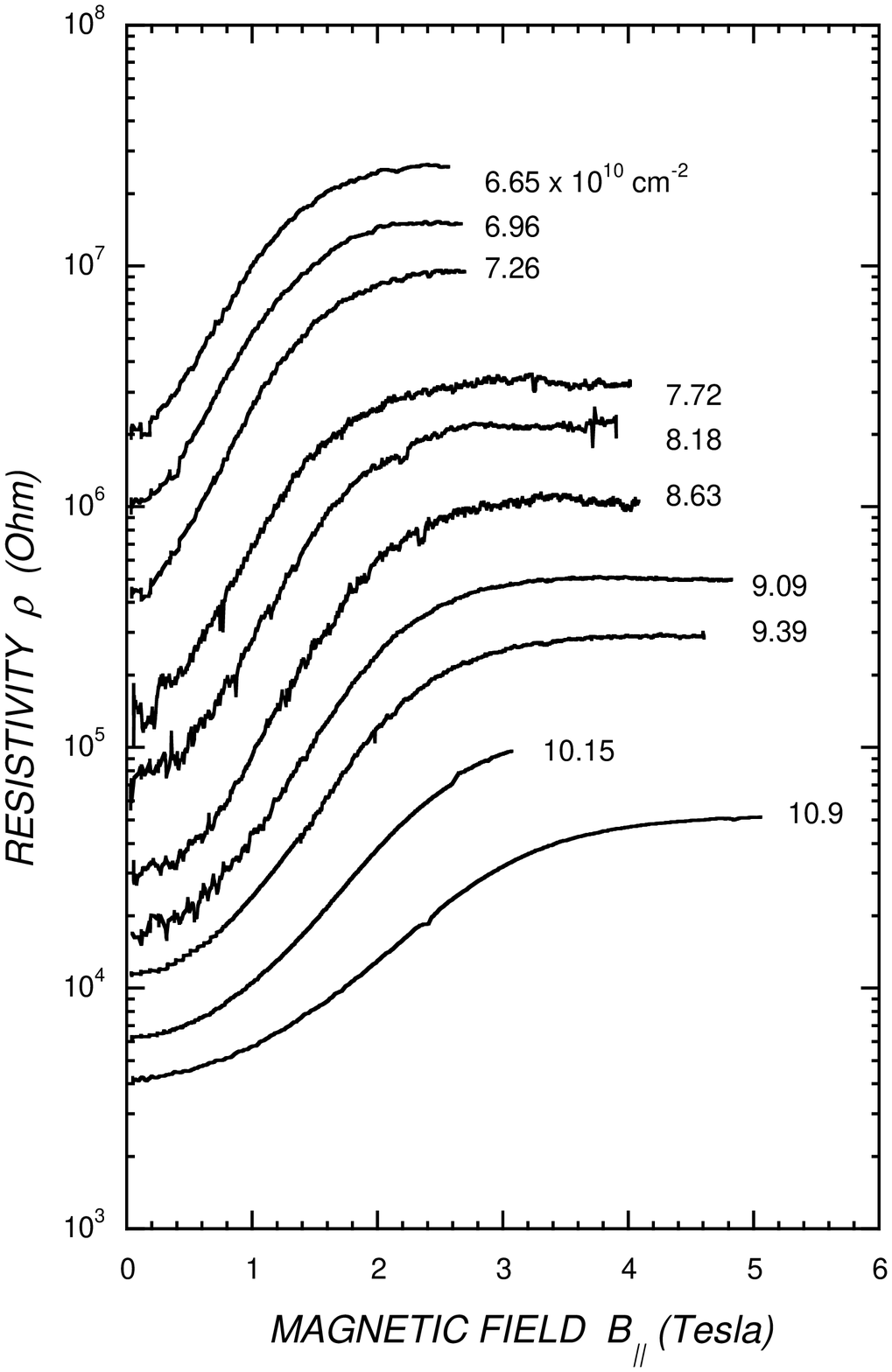,width=2.8in,bbllx=.5in,bblly=1.25in,bburx=7.25in,bbury=9.5in,angle=0}
}
\vspace{0.35in}
\hbox{
\hspace{-0.15in}
\refstepcounter{figure}
\parbox[b]{3.4in}{\baselineskip=12pt \egtrm FIG.~\thefigure.
Resistivity as a function of magnetic field applied parallel to the plane of a silicon MOSFET.  Data are shown for electron densities spanning the zero-field metal-insulator transition.  From Mertes~{\em et al.} (1999).\vspace{0.10in}
}
\label{fig:rh}
}
}

\subsection{The effect of a magnetic field}

Another unusual property of dilute two-dimensional systems is their enormous response to an external magnetic field.  We first consider the effect of a field $H_{||}$ applied parallel to the 2D plane.  A parallel field couples only to electron spins and not to their orbital motion (provided the fields are not so high that the magnetic length becomes comparable to the thickness of the 2D system).  The first observation of a suppression of conductivity in low-density Si MOSFETs by $H_{||}$ was reported by Dolgopolov~{\em et al.} in 1992.  The effect of a parallel magnetic field on $\rho(T)$ was studied in detail by Simonian~{\em et~al.} (1997b) and Pudalov~{\em et~al.} (1997a), also in Si MOSFETs.  In Fig.~\ref{fig:rh}, the resistivity is shown on a logarithmic scale as a function of parallel magnetic field at a fixed temperature of 0.3~K for several different electron densities on both sides of the metal-insulator transition.  The resistivity increases sharply as the magnetic field is raised, changing by more than an order of magnitude (up to four orders of magnitude at lower temperatures; see below).  It saturates above some density-dependent magnetic field $B_{sat}$ on the order of a few Tesla, remaining approximately constant up to the highest measuring field, $B_{||}=12$~Tesla (Pudalov~{\em et~al.}, 1997a).  In Si MOSFETs,
\vbox{
\vspace{-6mm}
\hbox{
\hspace{-1mm}
\psfig{file=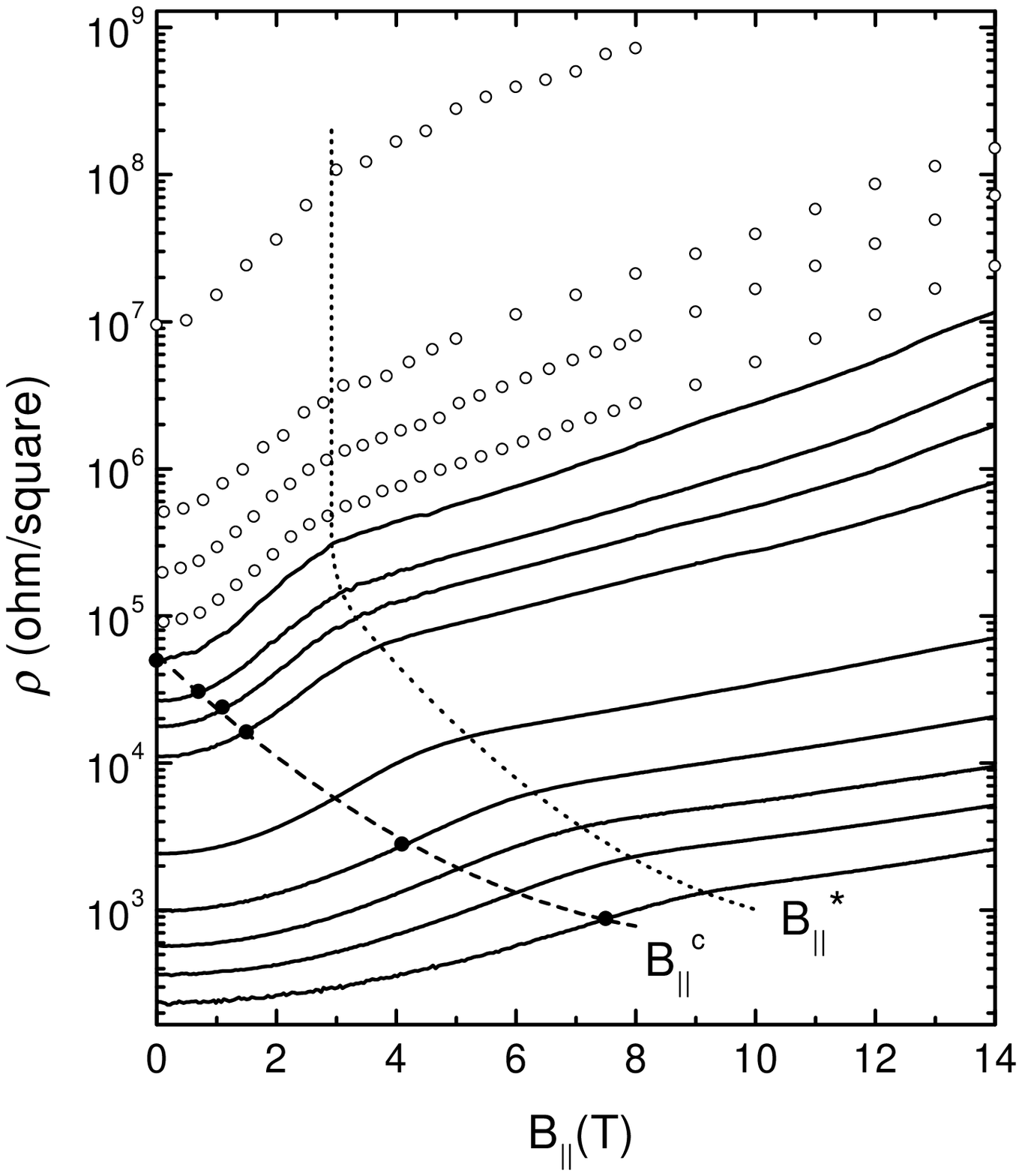,width=3.4in,bbllx=.5in,bblly=1.25in,bburx=7.25in,bbury=9.5in,angle=0}
}
\vspace{-0.35in}
\hbox{
\hspace{-0.15in}
\refstepcounter{figure}
\parbox[b]{3.4in}{\baselineskip=12pt \egtrm FIG.~\thefigure.
Resistivity as a function of $B_{||}$ in a p-GaAs/AlGaAs heterostructure at 50~mK at the following hole densities, from the bottom: 4.11, 3.23, 2.67, 2.12, 1.63, 1.10, 0.98, 0.89, 0.83, 0.79, 0.75, 0.67$\times$10$^{10}$~cm$^{-2}$.  The solid lines are for hole densities above $p_c$ and the open circles are for densities below $p_c$.  The solid circles denote the experimentally determined critical magnetic fields, and the dashed line is a guide to the eye.  $B^*_{||}$, the boundary separating the high and the low field regions, is marked as a dotted line.  From Yoon~{\it et al.} (2000).\vspace{0.10in}
}
\label{fig:yoon}
}
}
the magnetoresistance is independent of whether the parallel magnetic field is directed along or perpendicular to the measuring current.

The effect of a parallel magnetic field is qualitatively similar in p-GaAs/AlGaAs heterostructures, as shown in Fig.~\ref{fig:yoon} (Simmons~{\em et al.}, 1998; Yoon~{\em et al.}, 2000).  However, $\rho(B_{||})$ does not saturate to a constant value as in Si MOSFETs, but instead continues to increase with increasing field, albeit at a considerably slower rate.  As in the case of Si MOSFETs, there is a distinct knee that serves as a demarcation between the behavior in low and high fields.  Unlike MOSFETs, the magnetoresistance in p-GaAs/AlGaAs heterostructures is anisotropic; it depends on the relative directions of the measuring current, magnetic field, and crystal orientation.  These effects were studied in detail by Papadakis~{\em et al.} (2000).  In this system, there may be a contribution to the magnetoresistance anisotropy depending upon the angle between the current and the magnetic field due to finite thickness of the 2D layer as proposed by Das~Sarma and Hwang (2000).

It is noteworthy that the parallel field magnetoresistance is qualitatively the same for carrier densities above and below the zero-field critical density, $n_c$, {\it i.e.}, 
\vbox{
\vspace{-27.0mm}
\hbox{
\hspace{-10.5mm}
\psfig{file=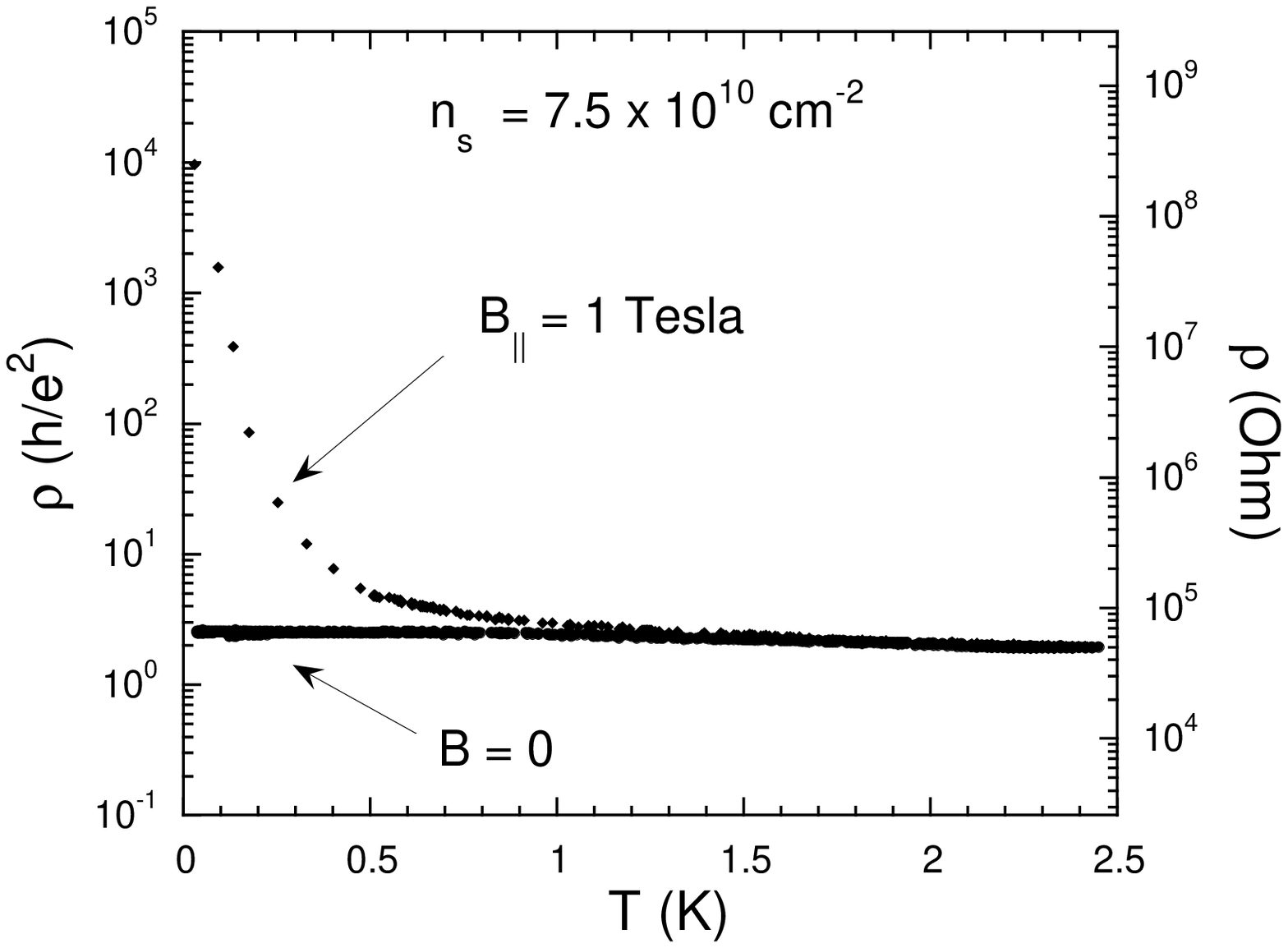,width=3.5in,bbllx=.5in,bblly=1.25in,bburx=7.25in,bbury=9.5in,angle=0}
}
\vspace{-0.45in}
\hbox{
\hspace{-0.15in}
\refstepcounter{figure}
\parbox[b]{3.4in}{\baselineskip=12pt \egtrm FIG.~\thefigure.
For the electron density corresponding to the critical density in zero field, the resistivity of a silicon MOSFET is shown as a function of temperature in zero field and in a parallel magnetic field of 1~Tesla.  Here $n_s=n_c=7.5 \times 10^{10}$ cm$^{-2}$; the measuring current is parallel to the field.  From Shashkin~{\em et al.} (2000).\vspace{0.15in}
}
\label{fig:sepH}
}
}
regardless of whether the temperature dependence is metallic or insulating in the absence of a magnetic field.  This suggests that the physical mechanism that gives rise to the magnetoresistance is the same in the two cases.  From an analysis of the positions of Shubnikov-de~Haas oscillations in tilted magnetic fields, Okamoto~{\em et~al.} (1999) argued that the magnetic field above which the resistivity saturates is the same as that required to fully polarize the electron spins.  A more direct demonstration of complete spin alignment for $B_{\parallel} \approx B_{sat}$ has recently been provided by small-angle Shubnikov-de Haas measurements of Vitkalov~{\em et al.} (2000).  Thus, the value of the resistance appears to be correlated with the degree of spin polarization of the 2D electron liquid.

As shown in detail below, a parallel magnetic field suppresses the metallic behavior and eventually turns the zero-field metal into a high-field insulator, for densities at least up to $1.5\,n_c$.  The extreme sensitivity to parallel field is dramatically illustrated for a high-mobility Si MOSFET in Fig.~\ref{fig:sepH}: at $T\approx30$~mK the resistance of the separatrix, which is temperature independent and near $2.5\,h/e^2$ in zero field, increases by almost four orders of magnitude and acquires an insulating temperature dependence in a parallel field of only 1~Tesla.  Thus, localized behavior, which appears to be absent at $n_s=n_c$ in $B=0$ even at resistances $\gtrsim h/e^2$, is ``restored'' in a magnetic field.  In contrast, the same field has a negligible effect on the resistance at temperatures higher than 1.5~K where the resistance can be described by the Drude formula.  We conclude that the enormous response observed at low temperatures is a consequence of effects other than parallel field-induced changes in carrier density or disorder strength.
\vbox{
\vspace{-1mm}
\hbox{
\hspace{3mm}
\psfig{file=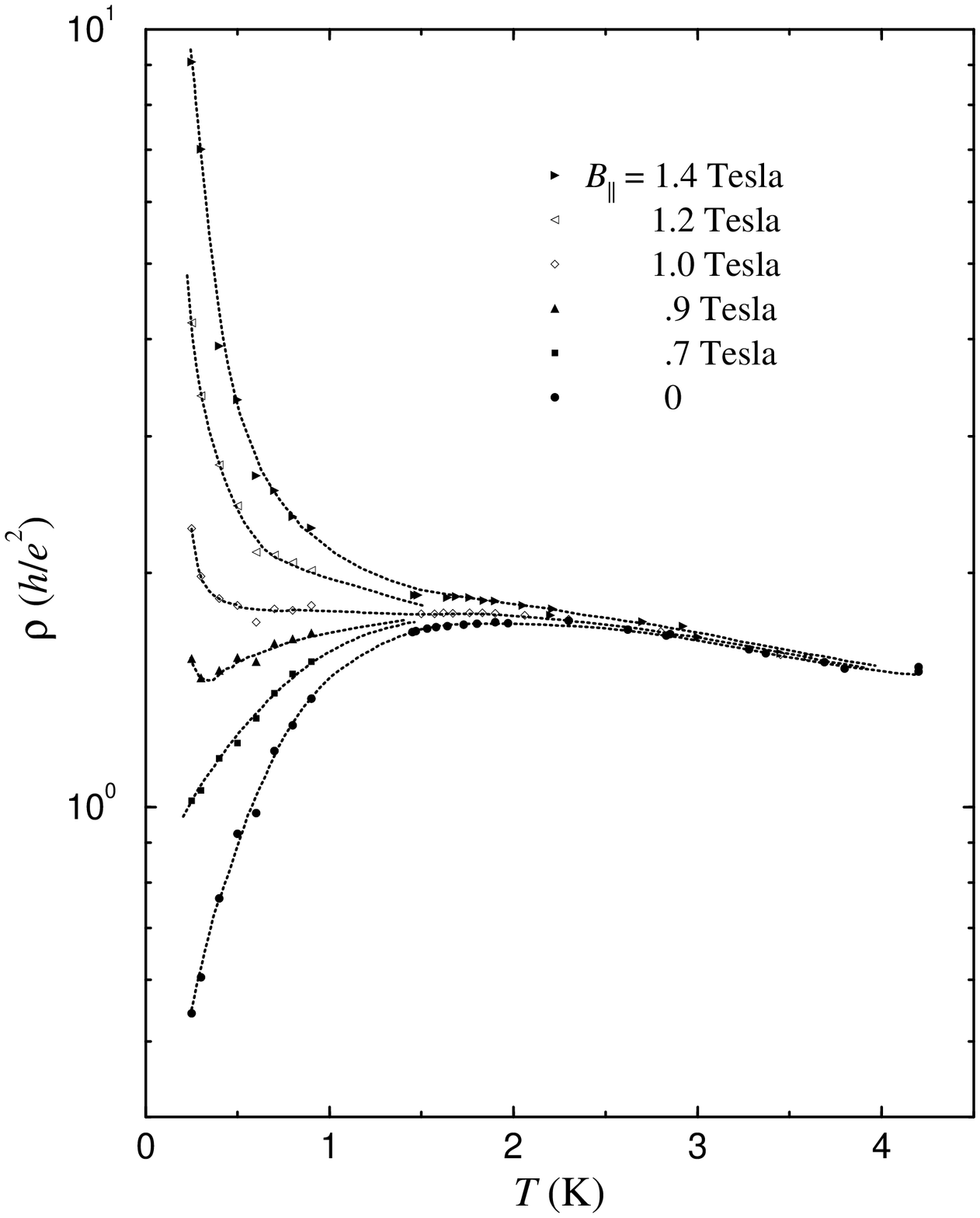,width=2.8in,bbllx=.5in,bblly=1.25in,bburx=7.25in,bbury=9.5in,angle=0}
}
\vspace{0.15in}
\hbox{
\hspace{-0.15in}
\refstepcounter{figure}
\parbox[b]{3.4in}{\baselineskip=12pt \egtrm FIG.~\thefigure.
Resistivity versus temperature for five different fixed magnetic fields applied parallel to the plane of a silicon MOSFET.  The electron density is $8.83\times10^{10}$~cm$^{-2}$.  From Simonian~{\em et al.}
(1997b).\vspace{0.10in}
}
\label{fig:rth}
}
}

Recent studies (Mertes~{\em et al.}, 2000) have shown that the resistance diverges somewhat more strongly with decreasing temperature in the high-field insulator (saturated region) than it does in the low-density, zero-field insulator; the resistance in high fields has the exponentially-activated variable range hopping form with an exponent greater than the value 1/2 (see Eq.~4) found in zero applied field.  This difference suggests that the spins play a role in determining the behavior in the insulating state.

Figure~\ref{fig:rth} shows how the temperature dependence of the resistance changes as the magnetic field is increased.  Here, the resistivity of a Si MOSFET with fixed density on the metallic side of the transition is plotted as a function of temperature in several fixed parallel magnetic fields between 0 and 1.4~Tesla.  The zero-field curve exhibits behavior typical for ``just-metallic'' electron densities: the resistivity is weakly-insulating at $T>T^*\approx2$~K and drops substantially as the temperature is decreased below $T^*$.  In a parallel magnetic field of only 1.4~Tesla (the upper curve), the metallic drop of the resistivity is completely suppressed, so that the system is now strongly insulating in the entire temperature range.  The effect of the field is negligible at temperatures above $T^*$, {\em i.e.}, above the temperature below which the metallic behavior in $B=0$ sets in.  Therefore, $T^*$ signals a temperature below which there is an abrupt onset of metallic behavior and below which the magnetoresistance becomes extremely large.  The Zeeman energy in a \vbox{
\vspace{-1mm}
\hbox{
\hspace{0mm}
\psfig{file=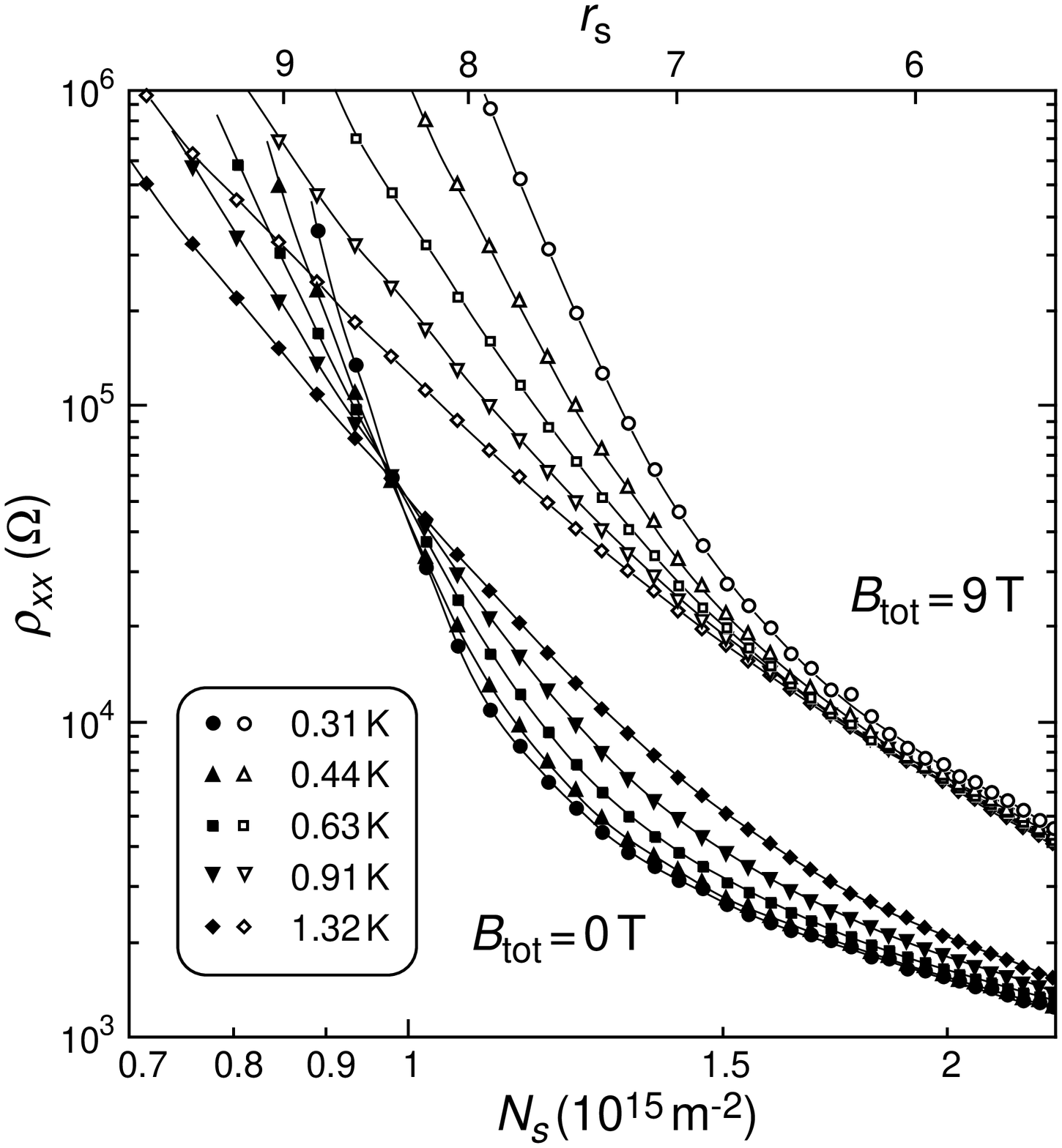,width=2.8in,bbllx=.5in,bblly=1.25in,bburx=7.25in,bbury=9.5in,angle=0}
}
\vspace{0.15in}
\hbox{
\hspace{-0.15in}
\refstepcounter{figure}
\parbox[b]{3.4in}{\baselineskip=12pt \egtrm FIG.~\thefigure.
Diagonal resistivity as a function of electron density for different temperatures in zero magnetic field (closed symbols) and in a parallel magnetic field of 9~Tesla (open symbols).  From Okamoto~{\em et~al.} (1999).\vspace{0.18in}
}
\label{fig:okamoto}
}
}
magnetic field strong enough to suppress the metallic behavior, $g\mu_BB_{sat}$, appears to be close to the thermal energy corresponding to $T^*$.  For the data shown in Fig.~\ref{fig:rth}, both $g\mu_BB_{sat}$ and $k_BT^*$ are about 0.25~meV.

As clearly demonstrated by the data shown in Fig.\ref{fig:okamoto}, taken by Okamoto {\em et al.} (1999) in Si MOSFETs, high parallel magnetic field eliminates the metal-insulator transition entirely.  In the absence of a magnetic field, $\rho(n_s)$ curves for different temperatures cross at a single point corresponding to $n_s=n_c$, where the resistivity is temperature-independent.  At electron densities above (below) the crossing point, the resistivity increases (decreases) with temperature, displaying metallic (insulating) behavior; also, there is an approximate symmetry of $\rho(n_s)$ curves about this point, as we discussed in Sec.~IIb.  In a parallel magnetic field of 9~Tesla, however, there is no crossing point, and there remains no sign of a metal-insulator transition.  Note that the effect of the magnetic field cannot be ascribed solely to a field-induced change in the critical electron density.

In Si MOSFETs, the temperature and electric field scaling and the resistivity-conductivity symmetry around the transition, which we described in the previous section, break down in the presence of even a weak ($\lesssim1$~Tesla) parallel magnetic field.  This feature prompted Simonian~{\em et~al.} (1997b) to conclude that the metallic behavior at $B=0$ is suppressed by an arbitrarily small magnetic field, and that there is no ``critical'' magnetic field.  On the other hand, recent data of Yoon~{\em et~al.} 
\vbox{
\vspace{-1mm}
\hbox{
\hspace{3mm}
\psfig{file=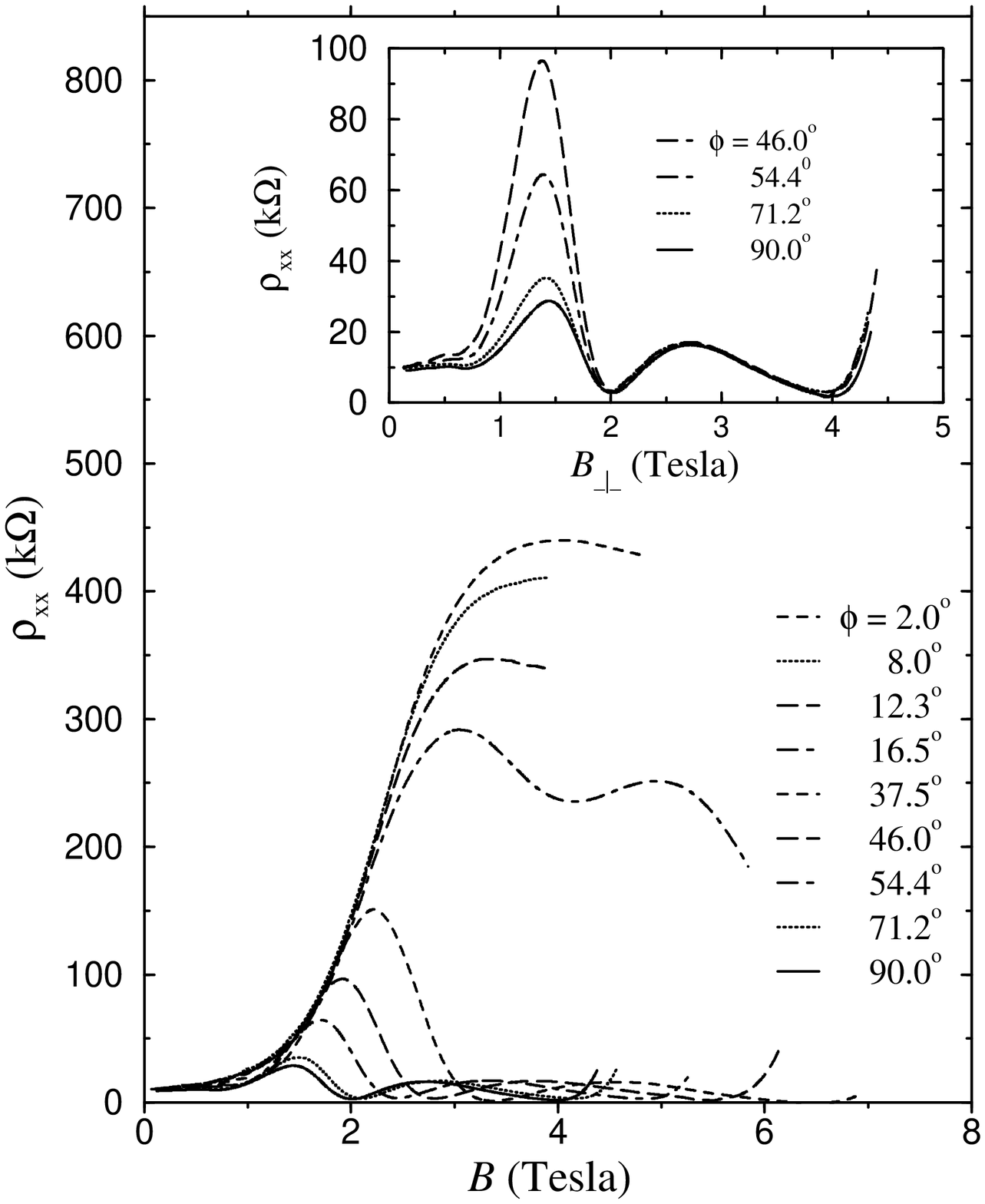,width=2.8in,bbllx=.5in,bblly=1.25in,bburx=7.25in,bbury=9.5in,angle=0}
}
\vspace{0.15in}
\hbox{
\hspace{-0.15in}
\refstepcounter{figure}
\parbox[b]{3.4in}{\baselineskip=12pt \egtrm FIG.~\thefigure.
Resistivity of a Si MOSFET at as a function of magnetic field applied at nine different angles $\phi$ with respect to the plane of the inversion layer.  Note that $\rho_{xx}$ deviates from the ``main'' curve at smaller magnetic fields as $\phi$ is increased.  The inset shows the resistivity of the same sample as a function of $B_{\perp}$ for four angles between the field and 2D plane.  $T=0.36$~K and $n_s=1.0\times10^{11}$~cm$^{-2}$.  From Kravchenko~{\em et al.} (1998).
\vspace{0.10in}
}
\label{fig:tilted}
}
}
(2000) suggest that there does exist a critical magnetic field in p-GaAs/AlGaAs heterostructures below which the metallic behavior survives: the $\rho(B_{||})$ curves taken at different temperatures cross at a single point so that the resistivity is independent of temperature at the field corresponding to the crossing point.  The difference relative to Si MOSFETs may derive from the fact that these measurements were performed on ultra-high mobility p-GaAs/AlGaAs heterostructures where the disorder potential may be much weaker than in MOSFETs.

In isotropic systems such as Si MOSFETs, studies have shown that the
metallic temperature dependence is suppressed in a similar way by magnetic fields applied at any angle relative to the 2D plane.
Fig.~\ref{fig:tilted} shows the longitudinal resistivity, $\rho_{xx}$, as a function of magnetic field applied at different angles.  For all angles the data follow approximately the same curve up to some value of magnetic field which depends on the tilt angle, above which orbital effects become dominant.  Above this field the resistivity traces out the standard quantum Hall effect (QHE) minima.  The resistivity deviates from the main curve at smaller magnetic fields as the angle between the field and the plane is increased. The larger perpendicular component causes stronger orbital effects which become dominant at a lower total field.  The magnetoresistance thus arises from the 
\vbox{
\vspace{23mm}
\hbox{
\hspace{-0.155in}
\psfig{file=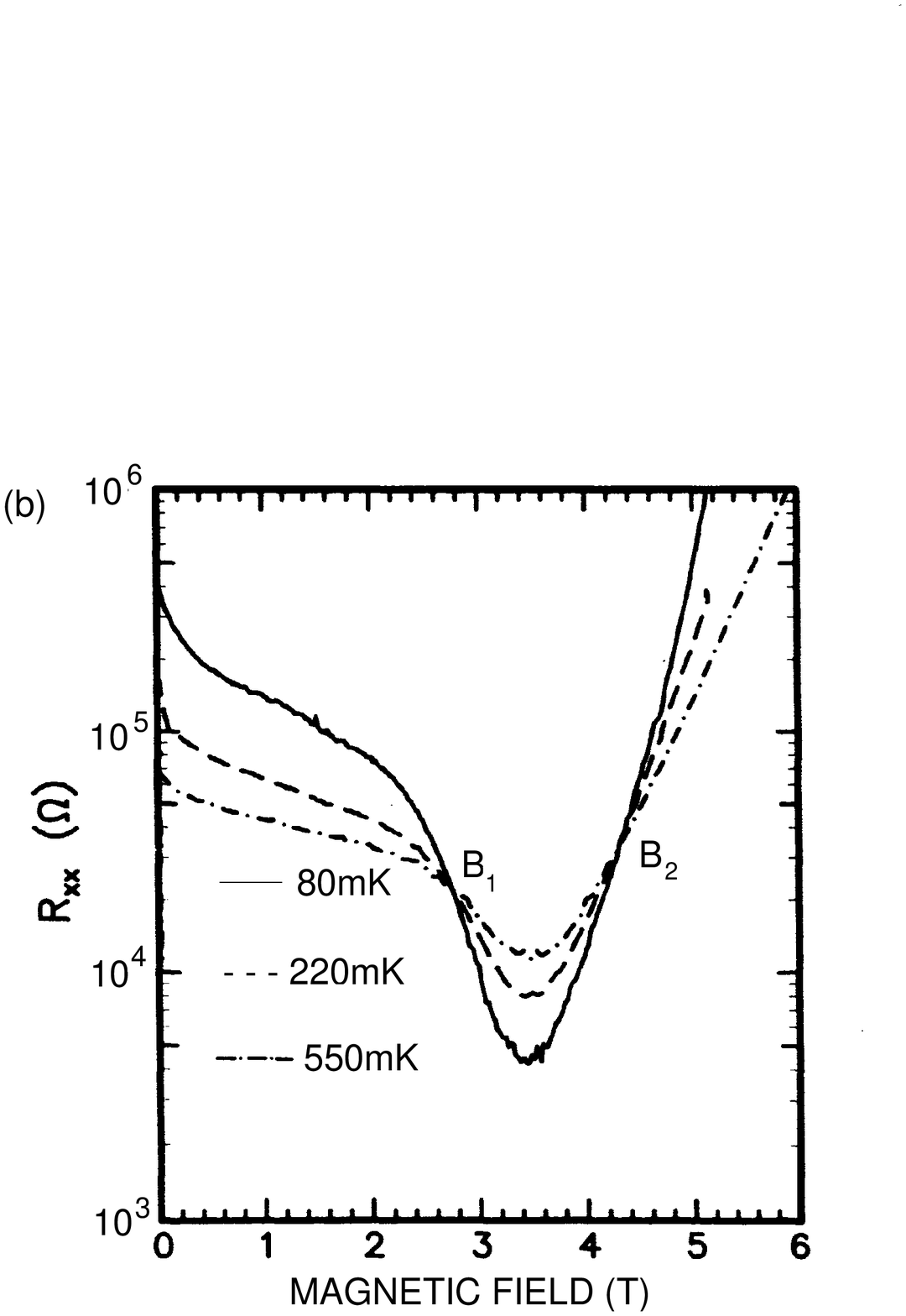,width=3.2in,bbllx=.5in,bblly=1.25in,bburx=7.25in,bbury=9.5in,angle=0}
}}\vspace{-57mm}
\vbox{
\vspace{-240mm}
\hbox{
\hspace{0.1in}
\psfig{file=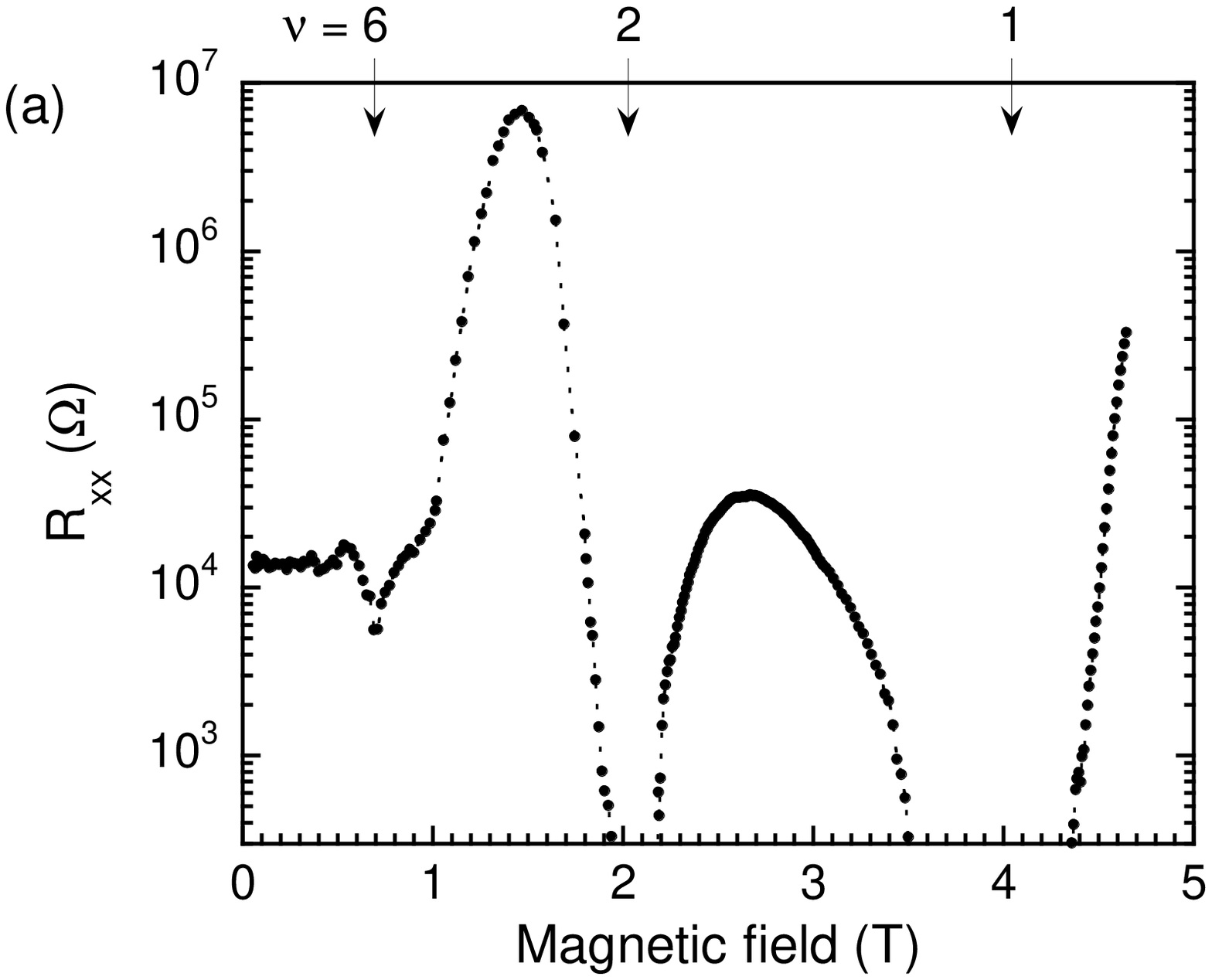,width=2.82in,bbllx=.5in,bblly=1.25in,bburx=7.25in,bbury=9.5in,angle=0}
}}\vspace{5.9cm}
\vbox{
\vspace{-3.8mm}
\hbox{
\hspace{7mm}
\psfig{file=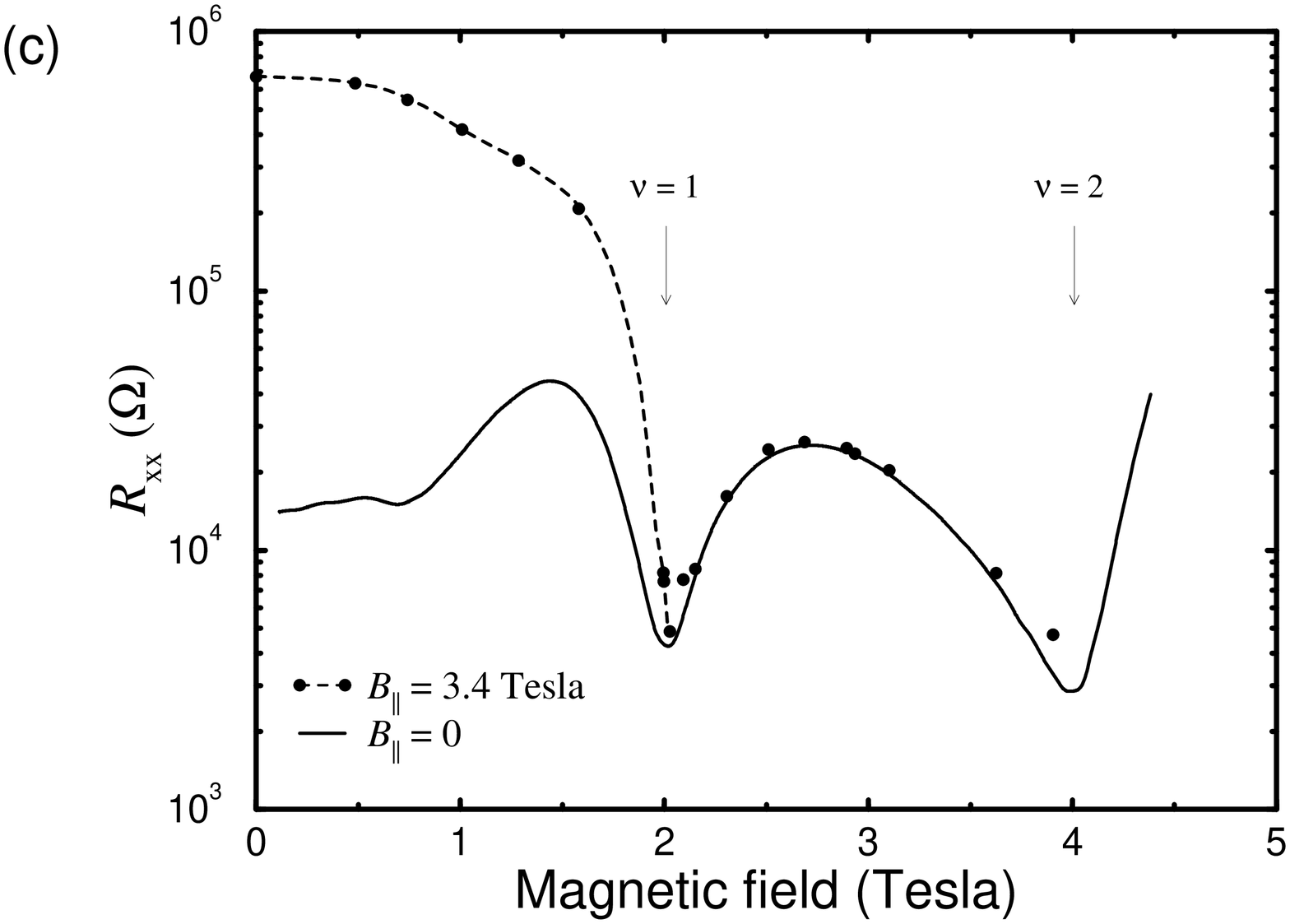,width=2.35in,bbllx=.5in,bblly=1.25in,bburx=7.25in,bbury=9.5in,angle=-90}
}
\vspace{0.3in}
\hbox{
\hspace{-0.15in}
\refstepcounter{figure}
\parbox[b]{3.4in}{\baselineskip=12pt \egtrm FIG.~\thefigure.
Longitudinal resistance as a function of perpendicular magnetic field, $B_{\perp}$.\\
(a) Si MOSFET at $T=35$~mK; $n_s=9.3\times10^{10}$~cm$^{-2}$.  Quantum Hall effect minima in the resistivity at filling factors $\nu=$1 and 2 and a Shubnikov-de~Haas minimum at $\nu=$6 are indicated by arrows.  From Pudalov~{\em et~al.} (1993b).\\
(b)~Low-mobility GaAs/AlGaAs heterostructure at three temperatures.  From Jiang~{\em et~al.} (1993).\\
(c)~Silicon MOSFET in the presence of a parallel field $B_{||}=3.4$~Tesla used to suppress the metallic behavior (solid symbols) and in zero magnetic field (solid line).  Temperature $T=0.36$~K and density $n_s=1.0\times10^{11}$~cm$^{-2}$.  From Kravchenko~{\em et al.} (1998).
}
\label{fig:perp}
}
}
superposition of two terms: the total field couples to the electron spins, yielding a large positive magnetoresistance, and the perpendicular field component couples to the orbital motion giving rise to the QHE.

Parallel field experiments in p-SiGe heterostructures yield results that are of particular interest.  This is an anisotropic system that is known to have very strong spin-orbit interactions.  Since an in-plane magnetic field cannot induce orbital motion perpendicular to the 2D plane, the field can couple to neither the orbit nor the spin, which are strongly coupled to each other.  Indeed, a parallel field was found to have negligible effect on the behavior of this system (Senz~{\em et~al.}, 1999; Coleridge~{\em et~al.}, 1999).  This provides further strong evidence that the giant magnetoresistance of Si MOSFETs and GaAs/AlGaAs heterostructures, where the spin-orbit coupling is weak, is due to coupling of magnetic field to the spin of the electrons (or holes).

We now consider the effect of a perpendicular magnetic field.  In very weak perpendicular fields, some systems display a small negative magnetoresistance characteristic of the suppression of weak localization; we shall return to this point later.  As first reported by D'Iorio~{\em et~al.} (1990), the magnetoresistance of Si MOSFETs displays anomalous behavior in a perpendicular magnetic field: as shown in Fig.\ref{fig:perp}(a), an initial, very large increase of the resistivity for $B_\perp\lesssim1$~Tesla is followed at higher field by the usual quantum Hall effect minima in the longitudinal resistance which occur when Landau levels are fully occupied, in this case at filling factors $\nu =$ 2 and 1 (here $\nu\equiv n_s\cdot ch/eB_\perp$).  The enormous positive magnetoresistance at $B_\perp\lesssim1.4$~Tesla is similar to that observed in parallel or tilted magnetic fields and is due to the suppression of metallic behavior.  The shape of $\rho(B_\perp)$ is especially intriguing because it is quite different from what has been observed in weakly-interacting, highly-disordered 2D systems, where re-entrant insulator-QHE-insulator transitions are found instead (Jiang~{\em et~al.}, 1993).  As shown in Fig.\ref{fig:perp}(b), these highly disordered systems are insulating at $B=0$, display a deep QHE resistance minimum in a field of a few Tesla, and then become insulating again.  This behavior is in agreement with the ``floating'' transition proposed by Khmelnitskii (1984) and Laughlin (1984) for non-interacting 2D systems, and with the global phase diagram of Kivelson, Lee, and Zhang (1992).  In this picture, the extended states, which exist at the center of each Landau level in high magnetic field, ``float up'' in energy as $B_\perp\rightarrow0$ so that their energies become infinitely high and the system is insulating at $B=0$.  This prediction was experimentally confirmed by Glozman~{\em et~al.} (1995) who showed that the extended states in highly-disordered n-GaAs/AlGaAs heterostructures do float up well beyond the Fermi energy as the field is decreased.

In the low-disordered, strongly-interacting dilute 
2D electron system in Si MOSFETs, however, the extended states deviate from their original positions at the centers of the Landau levels as $B_{\perp}\rightarrow0$, but rather than tending toward infinity, they coalesce at the Fermi level and remain at finite energy down to the lowest measured fields (Pudalov~{\em et~al.}, 1993b; Shashkin~{\em et~al.}, 1993, 1994a; Kravchenko~{\em et~al.}, 1995b).  This observation was interpreted (Shashkin~{\em et~al.}, 1993, 1994a) as evidence for the existence of a $B=0$ metallic state in this system.  Analogous experiments have recently been done by Dultz~{\em et~al.} (1998) and Hanein~{\em et~al.} (1999) in dilute p-GaAs/AlGaAs heterostructures with the same conclusion.  In the latter paper, it was shown that the insulator-QHE transition (corresponding to point $B_1$ in Fig.\ref{fig:perp}(b)) shifts to lower magnetic field as the hole density is increased, evolving gradually and continuously to the zero-field metal-insulator transition at $p_s=p_c$, the critical resistivities being the same for the finite-field insulator-QHE transition and zero-field metal-insulator transition.

Since the application of a parallel magnetic field suppresses the metallic behavior, it is reasonable to expect that in the presence of a fixed component of $B_{||}\sim$~few Tesla, the extended states should once again ``float up'' to large values, and the usual $\rho(B_\perp)$ behavior would then be restored. This is indeed what is observed: Fig.\ref{fig:perp}(c) shows $\rho_{xx}$ of a high-mobility Si MOSFET plotted as a function of perpendicular field in the presence of a parallel field $B_{||}=3.4$~Tesla.  The system now displays insulator-QHE-insulator transitions similar to those observed in highly-disordered n-GaAs/AlGaAs heterostructure by Jiang~{\em et~al.} (1993).\footnote{Note that the changes in resistance are less dramatic in Fig.\ref{fig:perp}(c) than in Fig.\ref{fig:perp}(a) because of the higher temperature (360~mK vs 35~mK).} In the presence of a parallel field large enough to suppress the metallic behavior, a strongly-interacting weakly-disordered 2D system behaves like a weakly-interacting strongly-disordered 2D system.

Note that at $B_{\perp}\gtrsim2$~Tesla, $\rho_{xx}(B_{\perp})$ follow approximately the same line regardless of the value of the parallel magnetic field component (see inset to Fig.\ref{fig:tilted}).  This suggests, once again, that the strength of the disorder and the electron density are not affected by the application of the parallel magnetic field even at low temperatures.  Similar behavior of $\rho_{xx}(B_{\perp})$ at $\nu<2$ regardless of the parallel magnetic field component was also reported by Okamoto~{\em et~al.} (1999).

Finally, we consider the influence of {\em weak} perpendicular magnetic fields.  In low-disordered noninteracting or weakly-interacting 2D systems, small perpendicular fields on the order of $0.05$~T are known to cause negative corrections to the resistance due to dephasing of the coherent backscattering process responsible for weak localization.  A weak negative magnetoresistance has been observed in small perpendicular magnetic fields in Si MOSFETs (see, {\it e.g.}, Brunthaler {\it et al.}, 1999) and p-GaAs/AlGaAs and p-SiGe heterostructures (Senz~{\em et~al.}, 2000; Simmons~{\em et~al.}, 2000; Coleridge~{\em et~al.}, 1999).  The existence of a negative magnetoresistance suggests that weak localization may still be present in these systems and may drive the system to a localized state at $T=0$.  However, the expected logarithmic dependence of the resistance on temperature due to weak localization is not observed at $B=0$ (except for at very high carrier densities) and appears to be overwhelmed by some other mechanism of unknown origin.  We note that the existence of a negative magnetoresistance does not necessarily imply carrier localization even within the framework of conventional localization theory with interactions, as discussed by Coleridge~{\em et~al.} (1999).  According to the expression for the corrections to the zero-field conductivity (Lee and Ramakrishnan, 1985),
\begin{equation}
\Delta\sigma(T)=(e^2/\pi h)(\alpha
p+1-\frac{3}{4}F^{\ast})\ln(kT\tau/\hbar)\label{Eq:LR},
\end{equation}
large enough values of the screening function $F^{\ast}$ lead to negative rather than positive $\Delta\sigma$ and thus to delocalization of the carriers as the temperature is reduced (here $p$ is the exponent describing the temperature dependence of the dephasing time $\tau_{\phi}\propto T^{-p}$, $\alpha$ is a constant of order 1, and $\tau$ is the elastic scattering time).  Finkelstein (1984) and Castellani~{\em et~al.} (1984, 1998) showed that $F^{\ast}$ can depend on temperature and have values much larger than 1.  Measurements of $F^{\ast}$ in a magnetic field have yielded values of 2.5 in p-SiGe (Coleridge~{\em et~al.}, 1999) and as high as 3.5 in Si MOSFETs (Bishop~{\em et al.}, 1982).  The issues of whether the metallic behavior survives to 0~K and whether it can be described within this theoretical framework are reviewed below in Sec.~III.

\subsection{Experiments other than transport}

Almost all measurements performed on these systems to date concern transport: the diagonal resistivity, the Hall resistivity and the magnetoresistance.  Very little is known regarding the thermodynamic behavior of 2D electron and hole systems.  Measurements of compressibility have recently been reported by two experimental groups.  Using a technique based on measurement of the capacitance, Dultz and Jiang (2000) found that the compressibility of 2D holes in p-GaAs/AlGaAs heterostructure changes sign at the critical density for the metal-insulator transition.  Ilani {\em et al.} (2000) determined the compressibility from measurements of the local chemical potential using single electron transistors.  They found qualitatively different behavior of the compressibility at low and high electron densities, with a crossover density that again agrees quantitatively with the transport critical density.  Both experiments suggest that the system undergoes a thermodynamic change at the transition.  The behavior of the compressibility is a key signature of the nature of the metal-insulator transition.  The freezing of the electron liquid into a disordered Wigner solid, for example, should be accompanied by a change of the compressibility from negative to positive.  This is discussed further in Sec.~III.

Fletcher {\em et al.} (2000) performed thermopower measurements in high-mobility silicon MOSFETs and found that the diffusion thermopower diverges at $n_s=n_c$ in a way similar to the divergence expected (Castellani~{\em et al.}, 1988) for a 3D Anderson metal-insulator transition, and consistent with the existence of a mobility edge in two dimensions.

\section{POSSIBLE EXPLANATIONS}
\label{sec:explanation}

The main observations which need to be explained are:
\begin{itemize}
\item Metallic behavior is displayed down to the lowest temperatures under conditions in which 2D systems are expected to show insulating behavior because of localization due to disorder (Anderson localization).
\item  The application of a magnetic field at an arbitrary angle to the plane of the two-dimensional electron liquid suppresses the metallic behavior and restores localization and other ``normal'' properties.
\end{itemize}

At present, there is no consensus about the nature of either of these effects. One thing that distinguishes the systems now under study from those examined in the past is that the interactions are enormous.  As explained in Sec.~II, the dimensionless measure of the interaction strength, $r_s$, is of order 10 or higher.  Thus we have an unambiguous example of the {\it strong-coupling many-body problem} for which theoretical methods are still poorly developed; it is a forefront area in theoretical condensed matter physics.  The old problem of the interplay of disorder (Anderson localization) and electron-electron interaction (Mott localization) is presented here in an extreme limit.  In spite of this situation, whether the electron-electron interaction plays the dominant role in the dynamics is controversial, as discussed below.

Various explanations have been suggested, ranging from non-Fermi liquid states and different kinds of superconductivity to single-particle physics based on  temperature-dependent scattering on charged traps and/or temperature-dependent screening. We now briefly review some of the models.

As mentioned in Sec.~I, the possibility that a metallic state can exist at zero magnetic field in two dimensions was first suggested by Finkelstein (1984) (see also Castellani~{\em et~al.}, 1984).  In this theory, the combined effects of interactions and disorder were studied by perturbative renormalization group (RG) methods. It was found that for a  weakly-disordered 2D system, an interaction parameter scales to infinitely large values --- thus out of the perturbative regime --- before zero temperature is reached.  Unfortunately the RG procedure fails as soon as this dimensionless coupling exceeds unity.  However, as the temperature is lowered, the resistivity first increases slightly and then begins to decrease (as often seen experimentally, see Sec.~IIB), just as the coupling becomes too large.  This suggests that a low-temperature metallic state might be achieved.  The theory does not contain a metal-insulator transition (in the absence of a magnetic field), nor is the nature of the possible metallic state revealed.  However, an external magnetic field, via Zeeman splitting, will drive the system back to the insulating state, in agreement with the experiments (see Sec.~IIF).  This scenario has not received general acceptance because the divergences which occur at non-zero temperature cause the theory to become uncontrolled.  It should also be remembered that the approach is perturbative and based on a Fermi liquid starting point.  In the present context, as discussed earlier, $r_s$ is so large that the theory's detailed applicability is in question.

Nevertheless, we may consider what happens within the RG scenario as the temperature is reduced further. The RG flow not only leads to a divergent interaction coupling but also to a divergent spin susceptibility.  This has been interpreted as signaling either the development of local moments (Finkelstein, 1984) or of ferromagnetism (Kirkpatrick and Belitz, 1996).  The latter is expected to occur at sufficiently large $r_s$ in the 2D interacting electron system (Ceperley and Alder, 1980). The onset of such time reversal breaking effects entails a crossover to a situation (a different universality class) in which the previously diverging coupling remains finite, even decreases, and eventually, at low temperature, an insulating state is once again obtained.  Thus, carried to its conclusion, the renormalization group description appears to indicate an intermediate region of metallic-like behavior but so far fails to produce a metallic state at zero temperature (Kirkpatrick and Belitz, 1999).  However, the details of the various possible behaviors have not yet been worked out.

The possibility of metallic behavior was reconsidered recently by Castellani~{\em et~al.} (1998). They argued that in the case of weak disorder, the theory remains under control in a wide temperature range if renormalization of the energy scale (relative to the length scale) is taken into account.  They found a quite complex temperature dependence of the resistivity which at low temperature crosses over to a ``metallic'' power-law temperature dependence and then remains finite at $T=0$.  This behavior does not correspond to the exponential dependence of Eq.\ref{Eq:exp}, which is observed in most of the experiments. At the same time, a positive magnetoresistance in a parallel magnetic field was predicted to be proportional to $(B/T)^2$, in agreement with both old and new experiments (Bishop, Dynes, and Tsui, 1982; Coleridge~{\em et al}, 1999).  Spin-flip scattering on magnetic impurities was also predicted to destroy the metallic state.  Recent experiments of Feng~{\em et~al.} (1999) showing a low-temperature crossover to an insulating state in relatively low-mobility Si MOSFETs were interpreted in this spirit.

This theory is capable of explaining some of the experimental observations on the metallic side (and not too close to $n_c$), at least qualitatively.  For further tests, experiments other than transport should be made (magnetic susceptibility, tunneling, {\it etc}).  It should be noted that strong temperature dependence of the Hall coefficient, predicted by this theory, is not seen in the experiment (Pudalov~{\em et~al.}, 1999b; Sarachik~{\em et~al.}, 2000).

Recently, Si and Varma (1998), building on the previous work, developed a theory which leads to a metal-insulator transition, without, however, giving a description of the metallic phase.  A feature of the earlier RG approach is that the compressibility (proportional to the inverse screening length $1/s$) is unrenormalized. This is valid at high enough density so that the screening is good and $s$ is less than the mean free path $\ell$.  Si and Varma pointed out that at the low electron densities (large $r_s$) of the experiments, one might expect $s>\ell$ and thus a renormalization of the compressibility.  They argued that for large $r_s$, the interaction becomes unscreened at low temperature and the compressibility approaches zero, as might be expected as one nears a metal-insulator transition.  Si and Varma calculated a large suppression of the conductivity which overcomes the weak increase found in the earlier RG analysis of Finkelstein (1984) and Castellani~{\em et~al.} (1984, 1998). This leads to a metal-insulator transition (at which both conductivity and compressibility go to zero) controlled by whether $s$ is larger or smaller than $\ell$.

As mentioned in Sec.~IIG, there are recent experimental determinations of the electron fluid compressibility (Dultz and Jiang, 2000, and Ilani~{\em et~al.}, 2000) in p-GaAs/AlGaAs heterostructures.  Both groups found the compressibility tending to zero near the metal insulator-transition.  This is what one expects if the insulating state is one in which the long-range Coulomb interaction becomes unscreened (Si and Varma, 1998), as would also be the case in a disordered Wigner solid (Chakravarty~{\em et~al.}, 1999).

A phenomenological approach to the problem was taken by Dobrosavljevi\'{c}~{\em et~al.} (1997), who argued that the existence of a metal-insulator transition in 2D violates no general scaling principles for interacting electrons.  They proposed a scaling analysis which showed that a disordered 2D system of interacting electrons should scale either to a perfect conductor or to an insulator in the limit of zero temperature. The analysis has a metal-insulator transition as a quantum critical point and gives the scaling collapse and the resistivity-conductivity symmetries near the metal-insulator transition which are observed in many experiments, as discussed in Sec.~IIB.  These authors pointed out that the metallic state is very unlikely to be a Fermi liquid since if the interactions were turned off the metallic behavior would disappear and the system would become an Anderson insulator.

Pursuing this theme, the effect of disorder on a model of a 2D non-Fermi liquid was discussed by Chakravarty~{\em et~al.} (1998).  While the origin of the non-Fermi liquid was not explained, they showed that for sufficiently strong interactions (which would occur at low density), a non-Fermi liquid state of interacting electrons is stable in the presence of disorder and is a perfect conductor (as conjectured by Dobrosavljevi\'{c}~{\em et~al.} 1997).  Otherwise, the disorder leads to localization as in the case of non-interacting electrons.

An approach which begins from the strong-interaction limit was taken by Chakravarty~{\em et~al.} (1999).  They took the point of view that the insulating state at $n_s<n_c$ is due to formation of a disordered Wigner solid (``Wigner glass").  A transition from insulator to metal at $n_s>n_c$ is due to the melting of this glass into a non-Fermi liquid state characterized by short range magnetic (singlet) correlations.  The latter might be stable against disorder as mentioned in the previous paragraph (Chakravarty~{\em et~al.}, 1998).  In the limit of high carrier densities, when the relative role of the interactions decreases, the system should again become weakly localized, in agreement with experiments.  For $n_s>n_c$, a magnetic field quenches the singlet correlations and is predicted to drive the system into an insulating state.

At sufficiently low densities, a Wigner crystal (or glass) is expected to form which will be pinned in the presence of even a small amount of disorder.  The experimental evidence for possible Wigner crystallization was reported several years ago by Pudalov~{\em et al.} (1993a) and more recently by Simmons~{\em et al.} (1998).  Based on transport studies in exceptionally clean p-GaAs/AlGaAs heterostructures, Yoon~{\em et al.} (1999) suggested that the insulating phase at $n_s<n_c$ is associated with the formation of a Wigner crystal rather than with single-particle localization.   However, Mills~{\em et al.} (1999) did not observe insulating behavior in p-GaAs/AlGaAs heterostructures at considerably lower densities, where the interactions are even stronger and $r_s$ larger.

A number of explanations have been based on the possibility of superconductivity in an interacting 2D electron gas (see, {\it e.g.}, Kelly and Hanke, 1981; Ren and Zhang, 1994; Belitz and Kirkpatrick, 1992, 1998; Phillips~{\em et al.}, 1998; Thakur and Neilson, 1998).  This scenario is tempting because of similarities between the metal-insulator transition in 2D and the ``superconductor-insulator" transition in thin metallic films (for a review see, {\it e.g.}, Goldman and Markovi\'{c}, 1998) and also because a magnetic field suppresses conducting states in both cases leading to saturation of the resistance.

As mentioned in Sec.~IIF, Yoon~{\em et al.} (2000) have found a critical parallel magnetic field for each ``metallic" density in very clean p-GaAs/AlGaAs heterostructures. The existence of such a critical field is consistent with what would occur for a superconductor, even in its fluctuation regime above $T_c$.  On the other hand, zero (or very low) resistance has never been observed in these 2D materials and there is no experimental evidence that pairing of carriers occurs.  Furthermore, the occurrence of a negative magnetoresistance in some samples (see Sec.~IIF) is counterindicative of superconductivity.

There have been several suggestions that some of the unusual behaviors observed in dilute 2D electron systems at low temperatures can be explained by mechanisms that are classical in nature.  Altshuler and Maslov (1999) proposed a mechanism for strong temperature and magnetic field dependence of the resistivity in Si MOSFETs based on charging/discharging of traps in the oxide close to the 2D layer (but see the criticism by Kravchenko~{\it et al.}, 1999b, and by Phillips, 1999).  Klapwijk and Das~Sarma (1999) proposed a scenario based on scattering of the electrons on charged ions at the oxide-semiconductor interface under conditions when the numbers of electrons and ions are comparable. They showed that this could lead to the very large magnetoresistance observed on the insulating side of the transition.  Subsequently, Das~Sarma and Hwang (2000) calculated $\rho(T)$ on the metallic side of the transition based on the assumption that $n_c$ carriers are frozen to interface impurities and the metal-insulator transition occurs when there are no free electrons left. By considering the temperature dependence of the screening of the scattering by impurities, they obtained a non-monotonic temperature dependence of resistivity similar to that found in experiments near $T^*$.  In this model, however, at lower temperature $\sim T_F\cdot(\rho\,e^2/h)$, the metallic resistive behavior saturates.  This is of the order of 10~K in Si MOSFETs close to the transition where $\rho\, e^2/h\gtrsim1$ and the Fermi temperature is $T_F\sim10$~K.  In the experiments, if there is saturation at all, it occurs at temperatures which are two orders of magnitude lower.

None of these semi-classical models explains why localization is absent in zero magnetic field, although it is conjectured by some of the authors that at lower temperatures the apparent metallic behavior for $n_s\gtrsim n_c$ (which is seen down to less than 1\% of the Fermi temperature) turns finally into a weak upturn of the resistance characteristic of weak localization in 2D.  In order to explain the temperature-independent separatrix observed in some experiments, these models would have to provide a mechanism that yields a temperature-dependent disorder which precisely cancels the Anderson localization, a coincidence which seems improbable.  Within these ``classical" scenarios, the electron-electron interaction has little effect in spite of its very large value.

A number of authors have suggested a percolation-type description of the metal-insulator transition.  He and Xie (1998) proposed such a transition in the two-dimensional electron system in Si MOSFETs at low electron densities, with percolation occurring between a conducting liquid phase separated by regions of an insulating vapor phase.  A percolation transition involving non-interacting electrons was developed by Meir (1999).  He considered the system to be inhomogeneous, consisting of electron (or hole) puddles connected by quantum point contacts.  The model is capable of explaining the metallic $\rho(T)$ for $n_s$ close to $n_c$, but it predicts the drop of resistance to be no more than a factor two in a degenerate system, while experimentally it is more than ten in Si MOSFETs.  Meir remarked that in a non-degenerate system, the drop might be arbitrarily large.  However, in the experiment, the dramatic drop of the resistance occurs only when the electron system is degenerate (Kravchenko {\em et al}., 1999a).  The drop is also substantial at larger $n_s$ where the inhomogeneities, if any, are weak.  In fact, all the percolation scenarios require the metallic phase to be inhomogeneous on some scale.  It would be of interest to examine this issue experimentally.  Recent results of Ilani {\em et al.} (2000) seem to indicate an inhomogeneous insulating phase, but a homogeneous metallic one.

There have been several numerical attempts to solve the problem of interacting electrons in the presence of disorder ({\em e.g.}, Pikus and Efros, 1994; Benenti {\em et al.}, 1999; Denteneer {\em et al.}, 1999; Shepelyansky and Song, 1999). The number of electrons that can be treated in these calculations is limited and they are sometimes restricted to the case of spinless electrons.  Nevertheless, it is of interest that the effect of interactions is to cause some delocalization of the electrons.

\section{CONCLUSIONS}

The recent availability of samples of exceptionally high mobility have 
made experiments possible in two-dimensional systems with very low densities of electrons or holes.  This has opened a new largely unexplored regime of strongly-interacting physics where the following novel properties have been reported:
\begin{itemize}
\item A strong metallic temperature dependence of the resistivity ($d\rho/dT>0$) is observed in clean dilute 2D systems at carrier densities above some critical value, while insulating ($d\rho/dT<0$) behavior is seen at densities below the critical value.  The metallic behavior starts at temperatures below some fraction of the Fermi temperature and continues down to the lowest accessed temperatures, $T/T_F<10^{-2}$.  At the critical density, there appears to be a transition from a metallic-like phase to a strongly localized one.  The latter phase is what is expected for 2D systems under the conditions where metallic behavior is seen in the present experiments.
\item Metallic behavior persists to rather high carrier densities (smaller $r_s$) but its relative strength decreases with density.  A weak insulating temperature dependence reminiscent of Anderson localization is observed at higher densities of the order of those used in the experiments in the 1980s.
\item An external magnetic field applied at arbitrary angle with respect to the 2D plane suppresses the metallic behavior and eliminates it completely at $n_s\lesssim1.5\,n_c$.  This is not due to a change in the level of disorder or in carrier density.  The suppression of the metallic behavior appears to be correlated with the degree of spin polarization.
\item These effects have been observed in five different electron and hole systems.  In some 2D systems, the resistivity was found to scale with temperature and/or electric field on both sides of a critical density, and a conductivity-resistivity symmetry was observed around the transition, consistent with a quantum critical point describing the zero temperature metal-insulator transition.  At the critical density, the resistivity below some temperature was found to be practically independent of temperature in the best samples.
\end{itemize}

A central question that must be answered by experiment is whether there is a true metal-insulator transition at $n_s=n_c$ and a metallic phase at intermediate densities between $n_c$ and the higher densities where localization is known to prevail.  This will require measurements of many properties other than transport.  For example, tunneling measurements, which yield information about the single-particle density of states, will shed light on whether these are Fermi liquids, and whether this is indeed a metal-insulator phase transition.  The magnetization of the system is a central aspect which must be investigated, particularly in light of all the evidence provided by transport measurements that the spins play a very important role (which they do in some theories).  Other measurements, which would provide valuable information, are the specific heat, ESR, and NMR.  Given the very small number of electrons contained in a thin 2D layer of material with electron densities $\approx 10^9$ to $\approx 10^{11}$ cm$^{-2}$, these will be very difficult experiments to perform.

The main theoretical issue is the description of the 2D electron (or hole) system in the neighborhood of the critical density. Should the existence of a metal-insulator transition be unambiguously confirmed experimentally, this will require a theory of the unusual metallic phase. The experiments already indicate that the metal would not be an ordinary one. The enormous parallel field magnetoresistance and field-induced shift of the metal-insulator transition are two striking unexpected features. The nature of the insulating phase and {\em its} large magnetoresistance also need to be understood.
        
The existence of a true insulator-metal transition at low density implies the possibility of a second metal-insulator transition at higher density to the regime of weak localization. The understanding of this requires that the theory of the metallic phase cover the range of densities between the two transitions.
        
Various descriptions have been proposed, ranging from the melting of a Wigner solid from the insulating side to the formation of one from the metallic side; from superconductivity to quantum percolation; from a semi-classical one-electron description with no metal-insulator transition to a non-Fermi liquid scenario.  While each of these is capable of explaining one or another part of the set of experimental observations, none of them provides a comprehensive picture.

\section*{ACKNOWLEDGMENTS}

E. Abrahams acknowledges support by NSF grants DMR-9632294 and DMR-9976665, S.~V. Kravchenko is supported by NSF grants DMR-9803440 and DMR-9988283 and Sloan Foundation, and M.~P. Sarachik acknowledges support by DOE grant No.\ DOE-FG02-84-ER45153 and partial support from NSF grant DMR-9803440.

\end{multicols}
\end{document}